\documentclass[12pt]{article}
\usepackage{amsmath, amssymb, a4}
\usepackage{hyperref}
\usepackage{graphics}
\newcommand{\car}{{\cal A}_r(\om)}
\newcommand{\hw}{\widehat{W}}
\newcommand{\wrar}{\rightharpoonup}

\newcommand {\fou}{{\cal F}}
\newcommand {\finv}{{\cal F}^{-1}}
\newcommand{\tb}{\textbf{b}}

 \newcommand{\nc}{\newcommand}
 \nc{\di}{\displaystyle} \nc{\ra}{\rightarrow}
 \nc{\al}{\alpha} \nc{\be}{\beta}\nc{\ve}{\varepsilon} \nc{\vp}{\varphi}
 \nc{\Ga}{\Gamma} \nc{\ga}{\gamma} \nc{\Om}{\Omega}
 \nc{\om}{\omega} \nc{\la}{\lambda} \nc{\Si}{\Sigma}
 \nc{\si}{\sigma} \nc{\Da}{\Delta} \nc{\da}{\delta}
 \nc{\na}{\nabla} \nc{\pa}{\partial} \nc{\ti}{\times}
 \nc{\N}{{\mathbb N}} \nc{\Z}{{\mathbb Z}} \nc{\C}{{\mathbb C}}
 \nc{\R}{{\mathbb R}} \nc{\V}{{\cal V}} \nc{\ek}{\protect\\[1ex]}
 \nc{\eq}[1]{\mbox{{(\ref{E#1})}}}
 \nc{\ome}{\Om^\ve} \nc{\les}{\lesssim}
  \nc{\lan}{\langle} \nc{\ran}{\rangle}
\nc{\nn}{\nonumber}
 \nc{\PP}{(\mathbb{BL})^i_{\la,\tx'}}
 \nc{\tx}{\texttt{x}} \nc{\ty}{\texttt{y}}
 \nc{\hty}{\hat{\ty}} \nc{\hb}{\hat{\beta}}
 \nc{\curl}{\text{curl}} \nc{\tbeta}{\tilde{\beta}}
 \nc{\tom}{\tilde{\om}} \nc{\ttp}{\tilde{\tilde{p}}}
 \nc{\ttvp}{\tilde{\tilde{\vp}}} \nc{\ha}{\frac{1}{2}} \nc{\s}{\tilde}
\nc{\qed}{\mbox{ }\nolinebreak\hfill \rule{2mm}{2mm}}

\renewcommand{\div}{{\rm div}\,}

\newtheorem{lem}{Lemma}[section]
\newtheorem{theo}[lem]{Theorem}
\newtheorem{propo}[lem]{Proposition}
\newtheorem{coro}[lem]{Corollary}
\newtheorem{rem}[lem]{Remark}
\newtheorem{tdefi}[lem]{Definition}
\numberwithin{equation}{section}

\begin{document}
\bibliographystyle{alpha}
\title{Maximal Regularity in Exponentially Weighted Lebesgue Spaces of
the Stokes Operator in Unbounded Cylinders}
\author{Myong-Hwan Ri and Reinhard Farwig}
\date{}
\maketitle
\begin{abstract}\noindent
We study resolvent estimates and maximal regularity of the Stokes operator in $L^q$-spaces with
exponential weights in the axial directions of unbounded cylinders of $\R^n,n\geq 3$.
For a straight cylinder we use exponential weights in the axial direction
and Muckenhoupt weights in the cross-section.
Next, for cylinders with several exits to infinity we prove that the Stokes
operator in $L^q$-spaces with exponential weights
generates an exponentially decaying analytic semigroup and has maximal regularity.

The proof for straight cylinders uses an operator-valued Fourier multiplier theorem and
unconditional Schauder decompositions based on the
${\mathcal R}$-boundedness of the family of solution operators
for a system in the cross-section of the cylinder parametrized by the phase variable
of the one-dimensional partial Fourier transform.
For general cylinders we use cut-off techniques based on the result
for straight cylinders and the case without exponential weight.

\end{abstract}
\section {Introduction}
\footnotetext[1] {Ri Myong-Hwan: Institute of Mathematics,
Academy of Sciences, DPR Korea,
email: {\tt math.inst@star-co.net.kp}}
\footnotetext[2] {Reinhard Farwig: Department of Mathematics,
Darmstadt University of Technology, 64289 Darmstadt, Germany,
email:  {\tt farwig@mathematik.tu-darmstadt.de}}

Let
\begin{equation}
\label{E1.2} \Om =\Om_0 \cup \bigcup_{i=1}^{m}\Om_i \subset \R^n,\; n\geq 3,
\end{equation}
be a cylindrical domain of $C^{1,1}$-class where $\Om_0$ is a
bounded domain and $\Om_i, i=1,\ldots, m,$ are disjoint semi-infinite
straight cylinders, that is, in possibly different coordinates,
$$ \Om_i=\{x^i=(x^i_1,\ldots, x^i_n)\in\R^n: \:x^i_n>0,\,
   (x^i_1,\ldots, x^i_{n-1})\in\Si^i\},$$
where the cross sections $\Si^i\subset \R^{n-1}$ are bounded domains
and $\Om_i\cap\Om_j=\emptyset$ for $i\neq j$.

Given a vector $\tb=(\be_1,\ldots,\be_m)$ with $\be_i\geq 0,
i=1,\ldots,m,$ and $1<q<\infty$ we introduce the space
\begin{equation}
\label{1.5}
\begin{array}{rl}
L^q_\tb(\Om) & \!\!= \{U\in L^q(\Om): e^{\be_i x^i_n}U|_{\Om_i}\in L^q(\Om_i), 1\leq i\leq m\},\ek
\|U\|_{L^q_\tb(\Om)} & \!\!= \big(\|U\|^q_{L^q(\Om_0)}
+\sum_{i=1}^m \|e^{\be_i x^i_n}U\|_{L^q(\Om_i)}^q \big)^{1/q}
\end{array}
\end{equation}
Moreover, let $W^{k,q}_\tb(\Om)$, $k\in\N$, be the space of functions
whose partial derivatives up to $k$-th order belong to $L^q_\tb(\Om)$,
where a norm is endowed in the standard way.
As a subspace we introduce $W^{1,q}_{0,\tb}(\Om)=\{u\in W^{1,q}_\tb(\Om):u|_{\pa\Om}=0\}$.
Let $L^q_\si(\Om)$ and $L^q_{\tb,\si}(\Om)$ be the completion of the set
$C^\infty_{0,\sigma}(\Om)=\{u \in C^\infty_0 (\Om)^n:\, \div u=0\}$
in the norm of $L^q(\Om)$ and $L^q_\tb(\Om)$, respectively.
Then we consider the Stokes operator
$A = A_{q,\tb} = -P_{q,\tb}\Da$ in $L^q_{\tb,\si}(\Om)$ with domain
\begin{equation}
\label{1.6}
{\cal D}(A_{q,\tb}) = W^{2,q}_\tb(\Om)^n\cap W^{1,q}_{0,\tb}(\Om)^n\cap L^q_{\tb,\si}(\Om),
\end{equation}
where $P_{q,\tb}$ is the Helmholtz projection of $L^q_\tb(\Om)$
onto $L^q_{\tb,\si}(\Om)$.

The goal of this paper is to study resolvent estimates and maximal $L^p$-regularity
of the Stokes operator in Lebesgue spaces with exponential weights in the axial direction.

There are many papers dealing with resolvent estimates
(\cite{FS94}, \cite{FS97}, \cite{Fr03},
\cite{Fr01}, \cite{Gi81}; see Introduction of \cite{FR05-1}
for more details) or maximal regularity
(see e.g. \cite{Ab04-1}, \cite{Fr02}, \cite{Fr01}) of Stokes operators
for domains with compact as well as noncompact boundaries.
General unbounded domains are considered in \cite{FKS} by replacing
the space $L^q$ by $L^q\cap L^2$ or $L^q+L^2$.
For resolvent estimates and maximal regularity in
unbounded cylinders without exponential weights in the axial direction
we refer to \cite{FR05-1}-\cite{FR05-4} and \cite{RY}.
For partial results in the Bloch space of
uniformly square integrable functions on a cylinder see \cite{Sc98}.


Despite of some references showing the existence of stationary
 flows in $L^q$-setting (e.g. \cite{NST99-1}-\cite{PT98})
and instationary flows in $L^2$-setting (e.g. \cite{Pi05}, \cite{PiZa08})
that converge as $|x|\ra \infty$ to some limit states (Poiseuille flow or zero flow)
 in unbounded cylinders,
resolvent estimates and maximal regularity of the Stokes operator in
$L^q$-spaces with exponential weights on unbounded cylinders do not
seem to be known yet.
\par\bigskip
The first main result of the paper concerns resolvent estimates and
maximal regularity of the Stokes operator in straight cylinders
$\Si\ti\R$; we get the result even in $L^q_\be(\R;L^r_\om(\Si))$,
$1<q,r<\infty$, with exponential weight $e^{\be x_n}$, $\be>0$, and
arbitrary Muckenhoupt weight $\om\in A_r(\R^{n-1})$ with respect to
$x'\in \Si$. We note that our resolvent estimate gives, in
particular when $\la=0$, a new result on the existence of a unique
flow with zero flux for the
 stationary Stokes system in $L^q_\be(\R,L^r_\om(\Si))$.

Next, for general cylinders $\Om$, we get resolvent estimates and
maximal $L^p$-regularity of the Stokes operator in $L^q_\tb(\Om)$,
$1<q<\infty$, using cut-off techniques.

The proofs for straight cylinders are mainly based on the theory of Fourier
analysis.
By the application of the partial Fourier transform along
the axis of the cylinder $\Si\ti\R$ the {\it generalized Stokes resolvent
system}
$$ \begin{array}{rcl}
\lambda U - \Da U +  \na P & =& F \quad \mbox{ in } \Si\ti\R,\ek
(R_\lambda) \hspace*{4cm} \div U & = & G \quad \mbox{ in }\Si\ti\R,\ek
u & = & 0 \quad \mbox{ on } \pa \Si\ti\R,
\end{array}$$
is reduced to the
{\it parametrized Stokes system} in the cross-section $\Sigma$:
$$ \begin{array}{rcll}
 (\lambda+ {\eta}^2 - \Da') \hat{U}' +  \na' \hat{P} & =& \hat{F}' \quad
 & \mbox{ in } \Sigma,\ek
(\lambda+ {\eta}^2 - \Da') \hat{U}_n +  i\eta \hat{P} & =& \hat{F}_n \quad
& \mbox{ in } \Sigma,\ek
(R_{\lambda,\eta}) \hspace{3cm} \div' \hat{U}'+i\eta\hat{U}_n & = &
   \hat{G} \quad &\mbox{  in }\Sigma,\ek
\hat{U}' = 0, \quad \hat{U}_n &=& 0 \quad &\mbox{ on } \pa   \Sigma,
\end{array}$$
which involves the Fourier phase variable $\eta \in {\mathbb C}$ as
parameter. Now, for fixed $\be\geq 0$ let
$$(\hat{u}, \hat{p}, \hat{f}, \hat{g})(\xi):=(\hat{U}, \hat{P}, \hat{F}, \hat{G})(\xi+i\be).$$
Then $(R_{\la,\eta})$ is reduced to the system
$$ \begin{array}{rcll}
 (\lambda+ (\xi+i\be)^2 - \Da') \hat{u}'(\xi) +  \na' \hat{p}(\xi) & =& \hat{f}'(\xi) \quad
 & \mbox{ in } \Sigma,\ek
(\lambda+ (\xi+i\be)^2 - \Da') \hat{u}_n(\xi) +  i(\xi+i\be) \hat{p}(\xi) & =& \hat{f}_n(\xi) \quad
& \mbox{ in } \Sigma,\ek
(R_{\lambda,\xi,\beta}) \hspace{3cm} \div' \hat{u}'(\xi)+i(\xi+i\be)\hat{u}_n(\xi) & = &
   \hat{g}(\xi) \quad &\mbox{  in }\Sigma,\ek
\hat{u}'(\xi) = 0, \quad \hat{u}_n(\xi) &=& 0 \quad &\mbox{ on } \pa   \Sigma.
\end{array}$$
We will get estimates of solutions to $(R_{\la,\xi,\be})$
independent of $\xi\in\R^*:=\R\setminus \{0\}$ and $\la$ in
$L^r$-spaces with Muckenhoupt weights, which yield
$\cal{R}$-boundedness of a family of solution operators
$a(\xi)$ for $(R_{\la,\xi,\be})$ with $g=0$ due to an extrapolation
property of operators defined on $L^r$-spaces with Muckenhoupt
weights. Then, an operator-valued Fourier multiplier theorem
(Theorem \ref{T4.6}) implies the estimate of $e^{\be x_n}U=\finv
(a(\xi) \fou f)$  for the solution $U$ to $(R_\lambda)$ with $G=0$
in the straight cylinder $\Si\ti\R$. In order to prove maximal
regularity of the Stokes operator in straight cylinders we use that
maximal regularity of an operator $A$ in a {\it UMD} space $X$ is
implied by the $\cal{R}$-boundedness of the operator family
\begin{equation}
\label{E1.1} \{\la(\la+A)^{-1}:\,\, \la\in {\rm i}\R\}
\end{equation}
in ${\cal L}(X)$, see \cite{We01}. Thus, the $\cal{R}$-boundedness
of \eq{1.1} for the Stokes operator $A:=A_{q,r;\be,\om}$ in
$L^q_\be(\R \!:\!L^r_\om(\Si))$ can be proved by virtue of Schauder
decomposition techniques.

The proofs for general cylinders, Theorem \ref{T2.4} and Theorem
\ref{T2.5}, use a cut-off technique
based on the result for resolvent estimates and maximal regularity
without exponential weights in \cite{FR05-4} and
the result (Theorem \ref{T2.3}) for straight cylinders.

\vspace{0.3cm}

This paper is organized as follows. In Section 2 the main results of
this paper (Theorem \ref{T2.1}, Corollary \ref{C2.2}, Theorem \ref{T2.3} --
 Theorem \ref{T2.5}) and
preliminaries are given. In Section 3 we obtain the estimate for $(R_{\lambda,\xi,\be})$ on
bounded domains $\Sigma\subset\R^{n-1}$, see Theorem \ref{T3.8}. In Section 4 proofs of the main
results are presented.

%
%

\section{Main Results and Preliminaries}
Let $\Sigma \ti {\mathbb R}$ be an infinite cylinder of ${\mathbb R}^n$
with bounded cross section $\Si\subset \R^{n-1}$ and with generic point
$x = (x', x_n) \in \Si\ti\R$. Similarly, differential operators in ${\mathbb R}^n$ and vector fields $u$ are split, in particular, $\Da=\Da'+\pa^2_n$, $\na=(\na', \pa_n)$, and $\div u = \div\hspace{-0.1cm}' u' + \pa_n u_n$.

For $q\in (1, \infty)$ we use the standard isomorphisms
$L^q(\Si\ti\R)\,\widetilde{=}\, L^q(\R;L^q(\Si))$ for classical Lebesgue spaces with norm
$\|\cdot\|_q=\|\cdot\|_{q;\Si\ti\R}$ and $W^{k,q}(\Si\ti\R)$, $k\in\N,$ for the
usual Sobolev spaces with norm $\|\cdot \|_{k, q;\Si\ti\R}$. We do not
distinguish between spaces of scalar functions and vector-valued
functions as long as no confusion arises. In particular, we use the
short notation $\|u, v\|_X$ for $\|u\|_X+\|v\|_X$, even if $u$ and
$v$ are tensors of different order.

Let $1<r<\infty$. A function $0\leq \om\in L^1_{\text{loc}}(\R^{n-1})$ is called
$A_r$-{\it weight} ({\it Muckenhoupt weight}) on $\R^{n-1}$ iff
$$ {\cal A}_r(\om):=\sup_Q \left(\frac{1}{|Q|}\int_Q\om\, dx'\right)
   \cdot\left(\frac{1}{|Q|}\int_Q\om^{{-1}/{(r-1)}}\, dx'\right)^{r-1}<\infty $$
where the supremum is taken over all cubes $Q\subset\R^{n-1}$
with edges parallel to the coordinate axes and $|Q|$ denotes the
$(n-1)$-dimensional Lebesgue measure of $Q$. We call ${\cal A}_r(\om)$ the
$A_r$-constant of $\om$ and denote the set of all $A_r$-weights on
$\R^{n-1}$ by $A_r=A_r(\R^{n-1})$. Note that
$$ \om\in A_r \quad \text{iff} \quad \om':=
   \om^{-1/{(r-1)}}\in A_{r'},\quad r'=r/{(r-1)},$$
and $A_{r'}(\om')=A_r(\om)^{r'/r}$.
A constant $C=C(\om)$ is called $A_r$-{\it consistent}
if for every $d>0$
$$\sup\,\{C(\om):\; \om\in A_r,\, {\cal A}_r(\om)<d\}<\infty.$$
We write $\om(Q)$ for $\int_Q \om\, dx'$.

Typical Muckenhoupt weights are the radial functions
$\om(x)=|x|^\al$: it is well-known that $\om \in A_r(\R^{n-1})$
if and only if $-(n-1)<\al<(r-1)(n-1)$;
the same bounds for $\al$ hold when $\om(x)=(1+|x|)^\al$ and
$\om(x)=|x|^\al(\log(e+|x|)^\beta$ for all $\beta\in\R.$
For further examples we refer to \cite{FS97}.

Given $\om\in A_r, r\in(1,\infty)$, and an arbitrary domain
$\Si\subset\R^{n-1}$ let
$$ L^r_\om(\Si)=\Big\{u\in L^1_{\text{loc}}(\bar{\Si}): \|u\|_{r,\om} =
   \|u\|_{r,\om;\Si} = \Big(\int_\Si |u|^r\om\,dx'\Big)^{1/r}<\infty\Big\}.$$
For short we will write $L^r_\om$ for $L^r_\om(\Si)$ provided that the
underlying domain $\Si$ is known from the context.
It is well-known that $L^r_\om$ is a separable reflexive Banach space with
dense subspace $C^\infty_0(\Si)$.
In particular, $(L^r_\om)^*=L^{r'}_{\om'}$. As usual, $W^{k,r}_\om(\Si)$,
$k\in\N$, denotes the weighted Sobolev space with norm
$$ \|u\|_{k,r,\om}=\Big(\sum_{|\alpha|\leq k}\|D^\alpha u\|^r_{r,\om}\Big)^{1/r},$$
where $|\alpha|=\alpha_1+\cdots+\alpha_{n-1}$ is the length of the
multi-index $\alpha=(\alpha_1,\ldots,\alpha_{n-1})\in\N^{n-1}_0$ and
$D^\alpha=\pa_1^{\alpha_1}\cdot\ldots\cdot \pa_{n-1}^{\alpha_{n-1}}$;
moreover,  $W^{k,r}_{0,\om}(\Si):=\overline{C^\infty_0(\Si)}^{\|\cdot\|_{k,r,\om}}$
and $W^{-k,r}_{0,\om}(\Si):=\big(W^{k,r'}_{0,\om'}(\Si)\big)^*$.
We also introduce the weighted homogeneous Sobolev space
$$ \hw^{1,r}_\om(\Si)=\left\{u\in {L^1_{\text{loc}}(\bar{\Si})}/\R:\:
   \na' u \in L^r_{\om}(\Si)\right\}$$
with norm $\|\na' u\|_{r,\om}$ and its dual space $\hw^{-1,r'}_{\om'}:=\big(\hw^{1,r}_\om\big)^*$
with norm $\|\cdot\|_{-1,r',\om'}= \|\cdot\|_{-1,r',\om';\Si}$.

Let $q,r\in (1,\infty)$. On an infinite cylinder $\Si\ti\R$, where $\Si$ is a
bounded $C^{1,1}$-domain of $\R^{n-1}$, we define the function space
$L^q(L^r_\om):=L^q(\R;L^r_\om(\Si))$ with norm
$$ \|u\|_{L^q(L^r_\om)}=\left(\int_\R \Big(\int_\Si |u(x',x_n)|^r
   \om(x')\,dx'\Big)^{q/r}\,dx_n\right)^{1/q}. $$
Furthermore, $W^{k;q,r}_\om(\Si\ti\R)$, $k\in\N,$ denotes the Banach space
of all functions in $\Si\ti\R$ whose partial derivatives of order up to $k$
belong to $L^q(L^r_\om)$ with norm
$\|u\|_{W^{k;q,r}_\om}=(\sum_{|\alpha| \leq k}
\|D^{\alpha}u\|^2_{L^q(L^r_\om)})^{1/2}$, where $\alpha\in\N^n_0$,
and let $W^{1;q,r}_{0,\om}(\Om)$ be the completion of the set
$C^\infty_0 (\Om)$ in $W^{1;q,r}_\om(\Om)$.
Given $\be\in\R$, let
$$ L^q_\be(L^r_\om):=\{u: e^{\be x_n}u\in L^q(L^r_\om)\}$$
equipped with the norm $\|e^{\be x_n}\cdot\|_{L^q(L^r_\om)}$, and for $k\in\N$ consider
$$ W^{k;q,r}_{\be,\om}(\Si\ti\R):=\{u: e^{\be x_n}u\in W^{k;q,r}_{\om}(\Si\ti\R)\}$$
with norm $\|e^{\be x_n}\cdot\|_{W^{k;q,r}_{\om}(\Si\ti\R)}$.
Finally, $L^q(L^r_\om)_\sigma$ and $L^q_\be(L^r_\om)_\sigma$ are completions
in the space $L^q(L^r_\om)$ and $L^q_\be(L^r_\om)$ of the set
$$ C^\infty_{0,\sigma}(\Si\ti\R)=\{u \in C^\infty_0 (\Si\ti\R)^n:\, \div u=0\},$$
respectively.

The Fourier transform in the variable $x_n$ is denoted by $\cal F$
or $\,\widehat{}\,$ and the inverse Fourier transform by ${\cal F}^{-1}$
or $^\vee$.  For $\varepsilon \in (0, \frac{\pi}{2})$ we define the
complex sector
$$ S_\varepsilon = \big\{ \lambda \in \C: \lambda\neq0, | \text{arg} \lambda |
   < \frac{\pi}{2}+ \varepsilon \big\}.$$

The first main theorem of this paper is as follows.
\begin{theo} {\bf (Weighted Resolvent Estimates)}
\label{T2.1}
Let $\Si\subset \R^{n-1}$ be a bounded domain of $C^{1,1}$-class with $\alpha_0>0$ and $\alpha_1>0$
being the least positive eigenvalue of the Dirichlet and Neumann Laplacian in $\Sigma$,
respectively, and let
$\bar\al:=\min\{\al_0,\al_1\}$, $\be\in (0,\sqrt{\bar\al})$,  $\al\in (0, \bar\al-\be^2)$, and
$0 <\varepsilon<\ve^*:=\arctan\big({\frac{1}{\be} \sqrt{\bar\al-\be^2-\al}}\big)$.
Moreover, let $1<q,r<\infty$ and $\om\in A_r.$

Then for every $F \in L^q_\be({\mathbb R}; L^r_\om(\Sigma))$, and
$\lambda\in-\alpha+S_\varepsilon$ there exists a unique solution
$(U,\na P)$ to $(R_\lambda)$ (with $G=0$) such that
$$ (\la+\al)U, \na^2U,\na P\in L^q_\be(L^r_\om)$$
and
\begin{equation}
\label{E2.1} \|(\lambda+\alpha) U, \na^2U, \na
P\|_{L^q_\be(L^r_\om)} \leq C \|F\|_{L^q_\be(L^r_\om)}
\end{equation}
with an $A_r$-consistent constant
$C=C(q,r,\alpha,\be,\ve,\Si,{\cal A}_r(\om))$ independent of $\lambda$.
\end{theo}

In particular, we obtain from Theorem 2.1
resolvent estimates of the Stokes operator in the cylinder
$\Si\ti\R$. Given the Helmholtz projection $P=P_{q,r;\be,\om}$ in
$L^q_\be(L^r_\om)$, see \cite{Fa03}, we define the Stokes operator
$A = A_{q,r;\be,\om}$ on $\Si\ti\R$ by $Au= -P \Da u$ for $u$ in the domain
\begin{equation}
\label{E2.2}
\mathcal D(A)= W^{2;q,r}_{\be,\om} ( \Si\ti\R) \cap W^{1;q,r}_{0,\be,\om} (\Si\ti\R)
 \cap L^q_\be(L^r_\om)_\sigma
   \subset L^q_\be(L^r_\om)_\sigma.
\end{equation}
%
%
\begin{coro} {\bf (Stokes Semigroup in Straight Cylinders)}
\label{C2.2}
Let $1<q,r<\infty$, $\om\in A_r(\R^{n-1})$,
$\varepsilon \in (0, \ve^*)$ and $\alpha\in(0,\bar\alpha-\be^2)$, $\be\in (0,\sqrt{\bar\al})$.

Then $-\alpha + S_\varepsilon$ is contained in the resolvent set of $-A=-A_{q,r;\be,\om}$,
and the estimate
\begin{equation}
\label{E2.3}
\|(\lambda + A)^{-1}\|_{{\cal L}(L^q_\beta(L^r_\om)_\sigma)}
\leq {\di {\frac{C}{|\lambda + \alpha|}}}, \quad
\forall \lambda \in -\alpha + S_\varepsilon,
\end{equation}
holds with an $A_r$-consistent constant $C=C(\Si,q,r,\al,\be,
\ve,\car)$.

As a consequence, the Stokes operator generates a bounded analytic semigroup
$\{e^{-tA};t\geq 0 \}$ on $L^q_\be(L^r_\om)_\sigma$ satisfying the estimate
\begin{equation}
\label{E2.4}
 \|e^{-tA}\|_{{\cal L}(L^q_\be(L^r_\om)_\sigma)}\leq C\, e^{-\alpha t}
 \quad \forall  t>0,
\end{equation}
with a constant $C=C(q,r,\al,\be,\ve,\Si,\car)$.
\end{coro}

The second important result of this paper is the {\it maximal regularity}
of the Stokes operator in an infinite straight cylinder.


\begin{theo} {\bf (Maximal Regularity in Straight Cylinders)}
\label{T2.3}
Let $1<p,q,r<\infty,$
$\om\in A_r(\R^{n-1})$ and $\be\in (0,\sqrt{\bar\al})$ be given.

Then the Stokes operator $A=A_{q,r;\be,\om}$ has maximal regularity in
$L^q_\be(L^r_\om)_\si$. To be more precise,
for each $F\in L^p(\R_+; L^q_\be(L^r_\om)_\si)$
the instationary problem
\begin{equation}
\label{E2.4n}
U_t+AU=F,\quad U(0)=0,
\end{equation}
has a unique solution
$U\in W^{1,p}(\R_+; L^q_\be(L^r_\om)_\si)\cap L^p(\R_+; \mathcal D(A))$
such that
\begin{equation}
\label{E2.5b} \|U, U_t, AU\|_{L^p(\R_+; L^q_\be(L^r_\om)_\si)}
   \leq C\|F\|_{L^p(\R_+; L^q_\be(L^r_\om)_\si)}.
\end{equation}

Analogously, for every $F\in L^p(\R_+; L^q_\be(L^r_\om))$,
the instationary Stokes system
\begin{equation*}
U_t-\Delta U+\na P=F,\quad\div U=0,\quad U(0)=0,
\end{equation*}
has a unique solution
$$(U,\na P)\in
\big(W^{1,p}(\R_+; L^q_\be(L^r_\om)_\si)\cap L^p(\R_+; \mathcal D(A))\big)
\times L^p(\R_+; L^q_\be(L^r_\om))$$
 satisfying the a priori estimate
\begin{equation}
\label{E2.5c}
\|U_t, U,\na U, \na^2U, \na P\|_{L^p(\R_+; L^q_\be(L^r_\om))}
   \leq C\|F\|_{L^p(\R_+; L^q_\be(L^r_\om))}
\end{equation}
with $C=C(\Si,q,r, \be, {\cal A}_r(\om))$.
Moreover, if
$e^{\al t} F\in L^p(\R_+; L^q_\be(L^r_\om))$ for some $\al\in(0,\bar\al-\be^2)$,
then the solution $u$ satisfies the estimate
\begin{equation}
\label{E2.5d}  \|e^{\al t}U, e^{\al t}U_t, e^{\al t}\na^2 U\|_{L^p(\R_+; L^q_\be(L^r_\om))}
   \leq C\|e^{\al t}F\|_{L^p(\R_+; L^q_\be(L^r_\om))}
\end{equation}
with $C=C(\Si,q,r,\al,\be,\car)$.
\end{theo}

\begin{rem}
The above statements for straight cylinders indeed hold for all
$\be\in (-\sqrt{\bar\al},\sqrt{\bar\al})$. This can be easily
checked by an inspection of the proofs as well as by an odd/even reflection argument introducing the new unknowns
$\tilde{u}(x',x_n)=(u'(x',-x_n), -u_n(x',-x_n)$, $\tilde{p}(x',x_n) = p(x',-x_n)$ and
by the result for $\be=0$ of \cite{FR05-3}.
\end{rem}

As a corollary of Theorem \ref{T2.3} we get the maximal regularity
result for general cylinders with several exits to infinity given by \eq{1.2}.
Recall the definition of the Stokes operator $A_{q,\tb} = -P_{q,\tb}\Delta$ and the spaces $L^q_{\tb}(\Om)$, $L^q_{\tb,\si}(\Om)$, see (\ref{1.5}), (\ref{1.6}). To the best of our knowledge, the existence of the Helmholtz projection $P_{q,\tb}$ has not been analyzed in the literature. However, in view of \cite{Fa03}, \cite{PT98}, \cite{Th02} it is clear that under the assumption $\beta_i\in (0,\sqrt{\overline \alpha_i})$, $i=1,\ldots,m$, where $\overline \alpha_i$ is the minimium of the smallest nontrivial eigenvalues of the Dirichlet and Neumann Laplacian in $\Si_i$, the Helmholtz projection $P_{q,\tb}: L^q_\tb(\Om) \to L^q_{\tb,\si}(\Omega)$ is a well-defined bounded operator.

\begin{theo} {\bf (Stokes Semigroup in General Cylinders)}
\label{T2.4}
Let $\Om\subset \R^n$ be a $C^{1,1}$-domain given by
\eq{1.2} and  let $\be_i>0$ satisfy the same
assumptions on $\be$ with $\Si^i$ in place of $\Si$. Then, the
Stokes operator $A_{q,\tb}(\Om)$ for $\tb=(\be_1,\ldots,\be_m)$
generates an exponentially decaying analytic semigroup
$\{e^{-tA_{q,\tb}}\}_{t\geq 0}$ in $L^q_{\tb,\si}(\Om)$.
 \end{theo}

\begin{theo} {\bf (Maximal Regularity in General Cylinders)}
\label{T2.5}
Under the general assumtions on $\Omega\subset \R^n$ and $\tb=(\be_1,\ldots,\be_m)$
as in Theorem \ref{T2.4} the Stokes operator $A_{q,\tb}$ has maximal regularity in
$L^q_{\tb,\si}(\Om)$. To be more precise, for any  $F\in L^p(\R_+;
L^q_{\tb,\si}(\Om))$, $1<p<\infty$, the Cauchy problem
\begin{equation}
\label{E2.8}
U_t+A_{q,\tb}U=F,\; U(0)=0, \quad \text{in } L^q_{\tb,\si}(\Om),
\end{equation}
has a unique solution $U$ such that
\begin{equation}
\label{E2.9}
\|U, U_t, A_{q,\tb}U\|_{L^p(\R_+;L^q_{\tb,\si}(\Om))}\leq C\|F\|_{L^p(\R_+;L^q_{\tb,\si}(\Om))}
\end{equation}
with some constant $C=C(p,q,\tb,\Om)$.

Equivalently, if $F\in L^p(\R_+; L^q_\tb(\Om))$,
then the instationary Stokes system
\begin{equation}
\label{E2.7}
\begin{array}{rcll}
U_t - \Da U +  \na P & =& F & \mbox{ in }\; \R_+\ti\Om,\ek
\div U & = & 0 & \mbox{ in }\; \R_+\ti\Om,\ek
U(0) & = & 0 & \mbox{ in }\; \Om,\ek
U & = & 0 & \mbox{ on }\; \pa \Om,
\end{array}
\end{equation}
has a unique solution $(U,\na P)\in L^p(\R_+; W^{2,q}_{\tb}(\Om)) \times
L^p(\R_+; L^q_\tb(\Om))$ such that $U_t\in L^p(\R_+; L^q_\tb(\Om))$ and
\begin{equation}
\label{E2.6}
\|U\|_{L^p(\R_+; W^{2,q}_\tb(\Om))} + \|U_t, \na P\|_{L^p(\R_+; L^q_\tb(\Om))}
   \leq C\|F\|_{L^p(\R_+; L^q_\tb(\Om))}.
\end{equation}
\end{theo}

\begin{rem}
\label{R2.5} {\rm  We note that in \eq{2.4n} and in \eq{2.8} we may take nonzero
initial values $u(0)=u_0$ in the interpolation space
$\big(L^q_\be(L^r_\om)_\si, \mathcal D(A_{q,r;\be,\om})\big)_{1-1/p,p}$
and
$U(0)=U_0\in \big(L^q_{\tb,\si}(\Om),\mathcal D(A_{q,\tb})\big)_{1-1/p,p}$,
respectively.

}
\end{rem}

For the proofs in Section 3 and Section 4, we need some preliminary results
for Muckenhoupt weights.
%
%
\begin{propo}{\rm (\cite[Lemma 2.4]{Fa03}}
\label{P2.5}
Let $1<r<\infty$, $\om\in A_r(\R^{n-1})$, and let $\Si\subset\R^{n-1}$ be a bounded domain. Then there exist $\tilde{s},s\in (1,\infty)$ satisfying

%
%
$$L^{\tilde{s}}(\Si)\hookrightarrow L^r_\om(\Si)\hookrightarrow L^s(\Si).$$
Here $\tilde{s}$ and $\frac{1}{s}$ are $A_r$-consistent. Moreover,
the embedding constants can be chosen uniformly on a set $W\subset A_r$ provided that
\begin{equation}
\label{E2.5}
\sup_{\om\in W}{\cal A}_r(\om)<\infty,\quad \om(Q)=1 \quad \text{for all  }\,\om\in W,
\end{equation}
for a cube $Q\subset \R^{n-1}$ with $\bar{\Si}\subset Q$.
\end{propo}
%
%
\begin{propo}{\rm (\cite[Proposition 2.5]{Fa03})}
\label{P2.6} Let $\Si\subset\R^{n-1}$ be a bounded Lipschitz domain
and let $1<r<\infty$.

(1) For every $\om\in A_r$ the continuous embedding
$W^{1,r}_\om(\Si)\hookrightarrow L^r_\om(\Si)$ is compact.

(2) Consider a sequence of weights $(\om_j)\subset A_r$ satisfying \eq{2.5} for
$W=\{\om_j:j\in\N\}$ and a fixed cube $Q\subset\R^{n-1}$ with $\bar{\Si}\subset Q$.
Further let $(u_j)$ be a sequence of functions on $\Si$ satisfying
$$  \sup_j \|u_j\|_{1,r,\om_j}<\infty \quad \text{and}\quad u_j\wrar 0\quad
    \text{in }W^{1,s}(\Si)$$
for $j\rightarrow \infty$ where $s$ is given by Proposition
\ref{P2.5}. Then
$$ \|u_j\|_{r,\om_j}\rightarrow 0 \quad \text{for   }j\rightarrow \vspace{0.2cm} \infty.$$

(3) Under the same assumptions on $(\om_j)\subset A_r$ as in (2) consider a
sequence of functions $(v_j)$ on $\Si$ satisfying
$$ \sup_j \|v_j\|_{r,\om_j}<\infty \quad \text{and}\quad v_j\wrar 0
   \quad \text{in }L^s(\Si)$$
for $j\rightarrow \infty$. Then considering $v_j$ as functionals on
$W^{1,r'}_{\om'_j}(\Si)$
$$ \|v_j\|_{(W^{1,r'}_{\om'_j}(\Si))^*}\rightarrow 0
   \quad \text{for   }j\rightarrow \infty.$$
\end{propo}
%
%
\begin{propo}
\label{P2.7}
Let $r\in(1,\infty),\; \om\in A_r$ and
$\Si\subset\R^{n-1}$ be a bounded Lipschitz domain. Then there exists
an $A_r$-consistent constant $c=c(r,\Si,{\cal A}_r(\om))>0$ such that
$$\|u\|_{r,\om}\leq c\|\na'u\|_{r,\om}$$
for all $u\in W^{1,r}_{0,\om}(\Si)$ and all $u\in W^{1,r}_\om(\Si)$ with vanishing integral mean $\int_\Si u\,dx'=0$.
\end{propo}
{\bf Proof:} See the proof of \cite[Corollary 2.1]{Fr01} and its conclusions;
checking the proof, one sees that the constant $c=c(r,\Si, {\cal A}_r(\om))$ is
$A_r$-consistent.
\hfill\qed
\medskip

Finally we cite the Fourier multiplier theorem in weighted spaces.
%
%
\begin{theo}  \label{T2.6} {\rm (\cite[Ch. IV, Theorem 3.9]{GR85})}
Let $m\in C^k(\R^k\setminus\{0\}),k\in\N,$ admit a constant $M\in \R$ such that
$$ |\eta|^\gamma |D^\gamma m(\eta)|\leq M
   \quad \text{for all}\quad \eta\in \R^k\setminus\{0\}$$
and multi-indices $\gamma\in\N^k_0$ with $|\gamma|\leq k$.
Then for all $1<r<\infty$ and $\om\in A_r(\R^k)$ the multiplier operator
$Tf={\cal F}^{-1}m(\cdot){\cal F}f$ defined for all rapidly decreasing functions
$f\in{\cal S}(\R^k)$ can be uniquely extended to a bounded linear operator from
$L^r_\om(\R^k)$ to $L^r_\om(\R^k)$. Moreover, there exists an $A_r$-consistent
constant $C=C(r,\car)$ such that
$$\|Tf\|_{r,\om}\leq CM\|f\|_{r,\om}\,,\quad f\in L^r_\om(\R^k)\,.$$
\end{theo}
%

\section{The problem $(R_{\la,\xi,\be})$ on the cross section}
In this section we get estimates for $(R_{\la,\xi,\be})$ independent
of $\la$ and $\xi\in\R^*$ in $L^r$-spaces on $\Si$ with Muckenhoupt
weights, where $\Si$ is a bounded $C^{1,1}$-domain of $\R^{n-1},
n\geq 3$.
 To this aim we rely partly on cut-off techniques using the
results for $(R_{\la,\xi})$ (i.e., the case $\be=0$)
in the whole and bent half spaces in \cite{FR05-3} (Theorem \ref{T3.1} below) and allow for a nonzero divergence $g$ in $(R_{\la,\xi,\be})$.
The main existence and uniqueness result in weighted $L^r$-spaces for $(R_{\la,\xi,\be})$
is described in Theorem \ref{T3.8}.

For the whole or bent half space $\Si$, $g\in \hw^{-1,r}_\om(\Si)+L^r_{\om}(\Si)$
and $\eta=\xi+i\be$ $\xi\in \R^*, \be\geq 0$,
we use the notation
$$ \|g; \hw^{-1,r}_\om+L^r_{\om,1/\eta}\| =
\inf\big\{\|g_0\|_{-1,r,\om}+\|g_1/\eta\|_{r,\om}:
       g=g_0+g_1, g_0\in \hw^{-1,r}_\om, g_1\in L^r_{\om}\big\}.$$
Note that obviously $W^{1,r}_\om(\Si) \subset \hw^{-1,r}_\om+L^r_{\om}$.
In the following we put $R_{\la,\xi}\equiv R_{\la,\xi,0}$ and, for simplicity, write $u$ for $\hat u$, $p$ for $\hat p$ etc..

\begin{theo}
\label{T3.1}
 Let $n \geq 3,\, 1<r<\infty,\, \om\in A_r(\R^{n-1})$,
$0<\ve <\frac {\pi}{2}$, $\xi \in {\mathbb R}^*$, $\la \in S_\ve$, $0< \ve <\pi /2$
and $\mu =|\la +\xi^2|^{1/2}$.
\begin{itemize}
\item[(i)] (\cite[Theorem 3.1]{FR05-3})
Let $\Si={\mathbb R}^{n-1}$.
If $f \in L^r_\om(\Si)$ and $ g \in W^{1,r}_\om(\Si)$, then the problem
$(R_{\la, \xi})$ has a unique solution $(u,p) \in W^{2,r}_\om(\Si) \ti W^{1,r}_\om(\Si)$
satisfying
\begin{equation}
\label{E3.9}
  \| \mu ^2u, \mu \na 'u, \na '^2u, \na 'p, \xi p\|_{r,\om}
 \leq c\big(\|f, \na 'g, \xi g\|_{r,\om}+\| \la g; \hw^{-1,r}_\om+L^r_{\om,1/\xi}\|\big)
\end{equation}
with an $A_r$-consistent constant $c=c(\ve,r,{\cal A}_r(\om))$  independent of
$\la$ and $\xi$.

\item[(ii)] (\cite[Theorem 3.5]{FR05-3}) Let $\Sigma=H_\sigma$ be a bent half space, i.e.
$$ \Si=H_\si=\{x'=(x_1,x'');\, x_1>\si(x''),x''\in {\mathbb R}^{n-2}\}$$
for a given function $\si \in C^{1,1}({\mathbb R}^{n-2})$.
Then there are $A_r$-consistent constants
$K_0=K_0(r,\ve,{\cal A}_r(\om))>0$ and $\la_0=\la_0(r,\ve,{\cal A}_r(\om))>0$ independent of
$\la$ and $\xi$ such that,
if $\| \na'\si \|_\infty \leq K_0$, for every
$f\in L^r_\om(\Si)$ and $ g\in W^{1,r}_\om(\Si)$
the problem
$(R_{\la,\xi})$ has a unique solution
$(u,p)\in (W^{2,r}_\om(\Si)\cap W^{1,r}_{0,\om}(\Si))\ti W^{1,r}_\om(\Si)$.
This solution satisfies the estimate
\begin{equation}
\label{E3.31}
\begin{array}{l}
\|\mu^2u,\mu\na'u,\na'^2u,\na'p,\xi p\|_{r,\om}\ek
\hspace{2cm}\leq c \big(\|f,\na'g,\xi g\|_{r,\om}+
                   \|\la g;\hw^{-1,r}_\om+L^r_{\om,1/\xi}\|\big)
\end{array}
\end{equation}
with an $A_r$-consistent constant $c=c(r,\ve,{\cal A}_r(\om))$.
\end{itemize}

\end{theo}

Now we turn our attention to  bounded domains $\Si\subset \R^{n-1}$ of $C^{1,1}$-class. Let $\alpha_0$ and $\al_1$
denote the smallest positive eigenvalues of the Dirichlet and Neumann Laplacian, respectively,
 i.e.,
\begin{equation}
\label{E3.10}
\begin{array}{l}
\alpha_0:=\inf \big\{\|\na u\|^2_2:\: u\in W^{1,2}_0(\Si), \|u\|_2=1\big\}>0,
\ek
\alpha_1:=\inf \big\{\|\na u\|^2_2:\: u\in W^{1,2}(\Si), \int_\Si u\,dx'=0,
\|u\|_2=1\big\}>0,\ek
\bar\al:=\min\{\al_0,\al_1\}.
\end{array}
\end{equation}

For fixed $\la\in \C\setminus (-\infty,-\alpha_0]$, $\eta=\xi+i\be$,
$\xi\in \R^*$, $\be\geq 0$,
 and $\om\in A_r$
we introduce the {\it parametrized Stokes operator} $S=S^\om_{r,\la,\eta}$ by
$$ S(u,p)= \left( \begin{array}{c}
           (\la+\eta^2-\Da')u'+\na'p\\
           (\la+\eta^2-\Da')u_n+i\eta p\\
           -\,\div\hspace{-0.1cm}_\eta u
\end{array} \right)$$
defined on ${\cal D}(S)={\cal D}(\Da'_{r,\om})\ti W^{1,r}_\om(\Si)$, where
${\cal D}(\Da'_{r,\om})=W^{2,r}_\om(\Si)\cap W^{1,r}_{0,\om}(\Si)$ and
$$\div\hspace{-0.1cm}_\eta u =\div'u'+i\eta u_n.$$
For $\om\equiv 1$ the operator $S^\om_{r,\la,\eta}$ will be denoted by $S_{r,\la,\eta}$.
Note that the image of ${\cal D}(\Da'_{r,\om})$ by $\div\hspace{-0.1cm}_{\eta}$
is included in $W^{1,r}_\om(\Si)$ and
$W^{1,r}_\om(\Si)\subset L^r_{0,\om}(\Si)+L^r_{\om}(\Si)$, where
$$ L^r_{0,\om}(\Si):=\big\{u\in L^r_\om(\Si):\: \int_{\Si}u\,dx'=0 \big\}.$$
Using Poincar\'{e}'s inequality in weighted spaces, see Proposition \ref{P2.7},
one easily gets the continuous embedding
$L^r_{0,\om}(\Si)\hookrightarrow \hw^{-1,r}_\om(\Si)$; more precisely,
$$ \|u\|_{-1,r,\om}\leq c\|u\|_{r,\om}\,, \quad u\in L^r_{0,\om}(\Si),$$
with an $A_r$-consistent constant $c>0$.
Moreover, we will use the notation
$$ \|g;L^r_{0,\om}+L^r_{\om,1/\eta}\|:=
   \inf \big\{\|g_0\|_{-1,r,\om}+\|g_1/\eta\|_{r,\om}:\:
   g=g_0+g_1, g_0 \in L^r_{0,\om}, g_1\in L^r_\om \big\},$$
the weighted Sobolev space $W^{1,r'}_{\om',\eta}$ on $\Si$ with norm
$\|\na' u,\eta u\|_{r',\om'}$ and its dual space denoted by $(W^{1,r'}_{\om',\eta})^*$.


First we consider the Hilbert space setting of $(R_{\la,\xi,\be})$.
For $\eta=\xi+i\be$, $\xi\in \R^*,$ $\beta\geq 0$, let us introduce
a closed subspace of $W^{1,2}_0(\Si)$ as
$$V_\eta:=\{u\in W^{1,2}_0(\Si):\div\hspace{-0.1cm}_\eta u=0\}.$$

\begin{lem}
\label{L3.1}
Let $\phi  
\in W^{-1,2}_0(\Si) = \big(W^{1,2}_0(\Si)\big)^*$ satisfy
$\lan \phi, v\ran_{W^{-1,2}_0(\Si), W^{1,2}_0(\Si)}=0$ for all $v\in V_\eta$.
Then, there is some $p\in L^2(\Si)$ with $\phi=(\na p, i\eta p)$.
\end{lem}
{\bf Proof:} This lemma can be proved in the same way as
\cite[Lemma 3.1]{FR05-1} with $\xi\in\R^*$ replaced by $\eta=\xi+i\be$.
\qed

\begin{lem}
\label{L3.2}
\begin{itemize}
\item[(i)] For any $g\in W^{1,2}(\Si)$, $\eta=\xi+i\be$, $\xi\in\R^*, \be\in\R$,
the equation $\div\hspace{-0.1cm}_\eta u=g$ has at least one solution
$u\in W^{2,2}(\Si)\cap W^{1,2}_0(\Si)$
and
$$ \|u\|_{2,2}\leq c\Big(\|g\|_{1,2}+\Big|\frac{1}{\eta}\int_\Si g\,dx'\Big|\Big),$$
where $c$ is independent of $g$.
\item[(ii)]
Let $\ve\in (0,\pi/2)$, $\be\in (0,\sqrt{\al_0})$ and
\begin{equation}
\label{E3.2}
\la\in \{-\al_0+\be^2+S_\ve\} \cap \big\{\la\in\C: \Re \la >
      -\frac{(\Im\la)^2}{4\be^2}-\al_0+\be^2\big\}.
\end{equation}
Then,
for any $f\in L^2(\Si)$, $g\in W^{1,2}(\Si)$
the system $(R_{\la,\xi,\be})$ has a unique solution
$(u,p)\in (W^{2,2}(\Si)\cap W^{1,2}_0(\Si)) \ti W^{1,2}(\Si)$.
\end{itemize}
\end{lem}
\begin{rem}
\label{R3.3} The assumption \eq{3.2} on $\la$ is satisfied for all
$\la\in -\al+S_\ve$ if either $\al\in (0, \al_0-\be^2)$ and $\ve\in
\big(0, \arctan( \frac{1}{\beta} \sqrt{\al_0-\be^2-\al}) \big)$ or
if $\al\in (0, \bar{\al}-\be^2)$ and
 $\ve\in \big(0, \arctan( \frac{1}{\beta} \sqrt{\bar{\al}-\be^2-\al}) \big)$.

\end{rem}
\par
\noindent
{\bf Proof of Lemma \ref{L3.2}:}  (i) Fix a scalar function $w\in
C^\infty_0(\Si)$ such that $\int_\Si w\,dx'=1$. Given $g\in
W^{1,2}(\Si)$, let $\bar{g}=\int_\Si g\,dx$ and consider the
divergence problem
$$\div'u'=g-\bar{g}w \quad\textrm{in } \Si, \quad u'|_{\pa\Si}=0,$$
which by \cite[Theorem 1.2]{FS94} has a solution $u'\in W^{2,2}(\Si)\cap W^{1,2}_0(\Si)$ with
$\|u'\|_{2,2}\leq c\|\na(g-\bar{g}w)\|_2\leq c\|g\|_{1,2}$.
Then $u:=(u',\frac{\bar{g}w}{i\eta})$ satisfies $\div\hspace{-0.1cm}_\eta u=g$ and the required estimate.

(ii)
By assertion (i), we may assume without loss of generality that
 $g\equiv 0$. Now, for fixed $\la\in -\al_0+\be^2+S_\ve$, define the
sesquilinear form $b: V_\eta\ti V_\eta\to \C$ by
 $$b(u,v):=\int_\Si \big((\la+\eta^2)u\cdot \bar{v}+\na'u\cdot\na'\bar{v}\big)\,dx'.$$
Obviously, $b$ is continuous in $V_\eta\ti V_\eta$. Moreover, $b$ is coercive, that is,
\begin{equation}
\label{E3.1}
|b(u,u)|\geq l(\la,\xi,\be)\|u\|^2_{1,2}
\end{equation}
with some $l(\la,\xi,\be)>0$.
In fact, let us write
\begin{equation}
\label{E3.1a}
b(u,u)=\int_\Si \big((\Re\la+\xi^2-\be^2\big)|u|^2 + |\na' u|^2)\,dx'
  +i\int_\Si (\Im\la+2\xi\be)|u|^2\,dx'
\end{equation}
and note that, due to the definition of $\al_0$,
$(\xi^2-\al_0)\|u\|_2^2+\|\na'u\|_2^2>0$ for all $\xi\in\R^*$ and $0\neq u\in V_\eta$.
Hence, if $\Re\la+\al_0-\be^2\geq 0$, then
$$|b(u,u)|\geq \Big|\int_\Si ((\Re\la+\xi^2-\be^2)|u|^2+|\na' u|^2)\,dx'\Big|
\geq (\xi^2-\al_0)\|u\|_2^2+\|\na'u\|_2^2,$$
where
$(\xi^2-\al_0)\|u\|_2^2+\|\na'u\|_2^2\geq \|\na'u\|_2^2$,
if $\xi^2-\al_0\geq 0$, and
$$ (\xi^2-\al_0)\|u\|_2^2+\|\na'u\|_2^2\geq \frac{\xi^2-\al_0}{\al_0} \|\na'u\|_2^2
   + \|\na'u\|_2^2 = \frac{\xi^2}{\al_0}\|\na'u\|_2^2$$
if $\xi^2-\al_0<0$.

Therefore, it remains to prove \eq{3.1} for the case  $\Re\la+\al_0-\be^2< 0$.

Note that if $\Im\la+2\xi\be\neq 0$ then $b(u,u) = b_{\la,\eta}(u,u)$ in \eq{3.1a}
coincides with $b_{\la_1,\xi}$ (on $V_\eta\times V_\eta$) where
$\la_1=\la-\be^2+2i\xi\be \in -\al_0+S_{\ve_1}$ with
$\ve_1=\max\big\{\ve, \arctan{\frac{|\Re\la+\al_0-\be^2|}{|\Im\la+2\xi\be|}} \big\}\in (0,\pi/2)$.
Hence, \eq{3.1} can be proved in the same way as \cite[Lemma 3.2 (ii)]{FR05-1}.

Finally suppose that
$$ \Im\la+2\xi\be= 0, \quad \textrm{i.e.,} \quad \xi=-\frac{\Im\la}{2\be}.$$
Since \eq{3.1} is trivial for the case $\Re\la+\xi^2-\be^2\geq 0$, we assume
that $\Re\la+\xi^2-\be^2< 0.$
In this case, note that due to the condition $\Re\la+\frac{(\Im\la)^2}{4\be^2}-\be^2>-\al_0$
there is some $c(\la,\be)>0$ such that
$$ 0>\Re\la+\frac{(\Im\la)^2}{4\be^2}-\be^2> c(\la,\be)-\al_0,  \quad c(\la,\be)-\al_0<0.$$
Then,
$$\begin{array}{rcl}
|b(u,u)|& \geq & \displaystyle\int_\Si \big((\Re\la
+\frac{(\Im\la)^2}{4\be^2}-\be^2)|u|^2+|\na' u|^2\big)\,dx'\\[2ex]
        & \geq &  \displaystyle\int_\Si \big(c(\la,\be)-\al_0)|u|^2+|\na' u|^2\big)\,dx'\\[2ex]
        & \geq &  \displaystyle\frac{c(\la,\be)}{\al_0}\|\na' u\|_2^2.
\end{array}$$
Now \eq{3.1} is completely proved.

By Lax-Milgram's lemma in view of \eq{3.1}, the variational problem
$$b(u,v)=\int_\Si f \cdot \bar{v}\,dx'\quad \forall v\in V_\eta,$$
has a unique solution $u$ in $V_\eta$. Then, by Lemma \ref{L3.1}, there is some $p\in L^2(\Si)$
such that
$$(\la+\eta^2-\Da')u'+\na' p=f', (\la+\eta^2-\Da')u_n+i\eta p=f_n.$$
Applying the well-known regularity theory for Stokes' system with nonzero divergence and Poisson's equation in $\Si$
to
$$-\Da'u'+\na' p=f'-(\la+\eta^2)u', \;\;\div'u' = -i\eta u_n,\;\; u'|_{\pa\Si}=0$$
and
$$-\Da'u_n=f_n-(\la+\eta^2)u_n-i\eta p, \;\; u_n|_{\pa\Si}=0,$$
respectively, we have $(u,p)\in (W^{2,2}(\Si)\cap W^{1,2}_0(\Si))\ti W^{1,2}(\Si)$.
\qed
\medskip

Now, we turn to considering $(R_{\la,\xi,\be})$ in spaces with
 weights with respect to cross sections as well.
%
%
\begin{lem}\
\label{L3.4}
Let $\be \in (0, \sqrt{\al_0})$, $\al\in (0,\al_0-\be^2)$,
 $\ve\in \big(0, \arctan( \frac{1}{\beta} \sqrt{\al_0-\be^2-\al}) \big)$, and
$\la\in -\alpha+S_\ve$. Moreover, fix $\xi\in\R^*$
 and $\om\in A_r$, $1<r<\infty$. Then the operator $S=S^\om_{r,\la,\eta}$ is
  injective and its range
is dense in $L^r_\om(\Si)\ti W^{1,r}_\om(\Si)$.
\end{lem}
\noindent {\bf Proof:} Since, by Proposition \ref{P2.5}, there is an
$s\in (1,r)$ such that $L^r_\om(\Si)\subset L^s(\Si)$, one sees
immediately that ${\cal D}(S^\om_{r,\la,\eta})\subset {\cal
D}(S_{s,\la,\eta}).$ Therefore, $S^\om_{r,\la,\eta}(u,p)=0$ for some
$(u,p)\in {\cal D}(S^\om_{r,\la,\eta})$ yields $(u,p)\in {\cal
D}(S_{s,\la,\eta})$ and $S_{s,\la,\eta}(u,p)=0$. Here note that
$S_{s,\la,\eta}(u,p)=0$ implies that
$$ S_{s,\la,\xi}(u,p) = \big((\be^2-2i\xi\be)u',(\be^2-2i\xi\be)u_n+\be p,\be u_n \big)^T. $$
Hence, by applying  \cite[Theorem 3.4]{FR05-1} a finite number of times and the Sobolev embedding
theorem, we get that $(u,p)\in (W^{2,2}(\Si)\cap W^{1,2}_0(\Si))
\ti W^{1,2}(\Si)$. Therefore, by Lemma \ref{L3.2} we obtain that $(u,p)=0$, i.e.,
$S^\om_{r,\la,\eta}$ is injective.

On the other hand, by Proposition \ref{P2.5}, there is an
$\tilde{s}\in (r,\infty)$ such that $S_{\tilde{s},\la,\eta}\subset
S^\om_{r,\la,\eta}$. Moreover, by Lemma \ref{L3.2}, for every $(f,
g)\in C^\infty_0(\Si)\ti C^\infty(\bar\Si)$, there is some $(u,p)\in
\mathcal D(S_{2,\la,\eta})$ with $S_{2,\la,\eta}(u,p)=(f,-g)$.
Applying the regularity result \cite[Theorem 1.2]{FS94} for the
Stokes resolvent system in $\Si$ a finite number of times  using the
Sobolev embedding theorem, it can be seen that $(u,p)\in \mathcal
D(S_{q,\la,\eta})$ for all $q\in (1,\infty)$, in particular, for
$q=\tilde{s}$. Therefore,
$$ C^\infty_0(\Si)\ti C^{\infty}(\bar\Si)
   \subset {\cal R}(S_{\tilde{s},\la,\eta})\subset{\cal R}(S^\om_{r,\la,\eta})
   \subset  L^r_\om(\Si)\ti W^{1,r}_\om(\Si), $$
which proves the assertion on the density of ${\cal R}(S)$.
\hfill\qed
\vspace{0.3cm}
\par
The following lemma gives a preliminary {\it a priori} estimate
for a solution $(u,p)$ of $S(u,p)=(f,-g)$.
%

%

\begin{lem}
\label{L3.5}
Under the assumptions on $r,\om,\al,\ve$ and $\be,\xi,\la$ as in Lemma \ref{L3.4}
 there exists an $A_r$-consistent constant $c=c(\ve, r,\al,\be,\Si,\car)>0$
such that for every $(u,p) \in {\cal D}(S_{r,\la,\eta}^\om)$,
\begin{equation}
\label{E4.1}
\begin{array}{l}
\hspace{-0.3cm} \|\mu^2_+u,\mu_+\na'u,\na'^2u,\na'p,\eta p\|_{r,\om}
    \leq c\big(\|f,\na'g,g,\xi g\|_{r,\om}+|\la|\|g;
    L^r_{0,\om}+ L^r_{\om,1/\eta}\| \\[1ex]
\hspace{56mm} +\|\na'u,\xi u,p\|_{r,\om} +
|\la|\| u\|_{(W^{1,r'}_{\om'})^*}\big);
\end{array}
\end{equation}
here $\mu_+=|\la+\alpha+\xi^2|^{1/2}, (f,-g)=S(u,p)$ and
$(W^{1,r'}_{\om'})^*$ denotes the dual space of $W^{1,r'}_{\om'}(\Si)$.
\end{lem}
{\bf Proof:} The proof is divided into two parts, i.e., the cases $\xi^2>\be^2$ and $\xi^2\leq \be^2$.

 The proof of the case $\xi^2>\be^2$
is based on a partition of unity in $\Si$ and on the
localization procedure reducing the problem to a finite number of
problems of type
$(R_{\la,\xi})$ in bent half spaces and in the whole space ${\mathbb R}^{n-1}$.
Since $\pa\Si\in C^{1,1}$, we can cover $\pa\Si$ by a
finite number of balls $B_j, j\geq 1$, such that, after a translation
and rotation of coordinates, $\Si\cap B_j$
locally coincides with a bent half space $\Si_j=H_{\si_j}$
where $\si_j\in C^{1,1}(\R^{n-1})$ has  compact support,
$\si_j(0)=0$ and $\na'' \si_j(0)=0$. Choosing the balls $B_j$
small enough (and its number large enough) we may assume that
$\|\na''\si_j\|_{\infty}\leq K_0(\ve,r,\Si,\car)$ for all $j\geq 1$ where $K_0$
was introduced in Theorem  \ref{T3.1} (ii).

According to the covering
$\pa\Si\subset \bigcup_j B_j$ there are non-negative cut-off functions
$\vp_j\in C^{\infty}({\mathbb R}^{n-1})$, $0\leq j\leq m$, such that
\begin{equation}
\label{E3.8}
 \sum\nolimits_{j=0}^m \varphi_j\equiv 1 \text{ in }\Si,
   \;\; \text{supp } \varphi_0\subset \Si,\;\; \text{supp}\, \varphi_j \subset B_j,\;\;\; j\geq 1.
\end{equation}

Given $(u,p)\in {\cal D}(S)$ and $(f,-g)=S(u,p)$, we get for each
$\varphi_j,\, j\geq 0,$ the local $(R_{\la,\xi})$-problems
\begin{equation}
\label{E4.2}
\begin{array}{rcl}
(\la+\xi^2-\Da')(\varphi_ju')+\na'(\varphi_jp)  & = & f'_j\ek
(\la+\xi^2-\Da')(\varphi_ju_n)+i\xi(\varphi_jp) & = & f_{jn}\ek
       \div\hspace{-0.1cm}_{\xi}(\varphi_ju)    & = & g_j
\end{array}
\end{equation}
for $(\varphi_ju, \varphi_j p), j\geq 0$, in ${\mathbb R}^{n-1}$ or $\Si_j$; here
\begin{equation}
\label{E4.3}
\begin{array}{rcl}
f'_j & = & \varphi_jf'-2\na'\varphi_j\cdot\na'u'-(\Da'\varphi_j)u'
+(\be^2-2i\xi\be) (\vp_j u')+(\na'\varphi_j)p\ek
f_{jn} & = & \varphi_jf_n-2\na'\varphi_j\cdot\na'u_n-(\Da'\varphi_j)u_n
+(\be^2-2i\xi\be) (\vp_j u_n)+\be(\vp_j p)\ek
g_j & = & \varphi_jg+\na'\varphi_j\cdot u'+\be\vp_j u_n.
\end{array}
\end{equation}

To control $f_j$ and $g_j$ note that $u=0$ on $\pa\Si$; hence Poincar\'{e}'s inequality
for Muckenhoupt weighted spaces (Proposition \ref{P2.7}) yields for all $j\geq 0$ the estimate
\begin{equation}
\label{E4.4}
\|f_j,\na'g_j,\xi g_j\|_{r,\om;\Si_j}\leq c(\|f,\na'g,g,\xi g\|_{r,\om;\Si} +
   \|\na'u,\xi u,p\|_{r,\om;\Si}),
\end{equation}
where $\Si_0\equiv \R^{n-1}$ and $c=c(\be)>0$ is $A_r$-consistent.

The crucial terms are the norms $\|g_j;\hw^{-1,r}_\om(\Si_j)+L^r_{\om,1/\xi}(\Si_j)\|$
which appear when Theorem \ref{T3.1} is applied to \eq{4.2}. For their analysis
let $g=g_0+g_1$ denote any splitting of $g\in L^r_{0,\om}+L^r_{\om,1/\eta}$.
Defining the characteristic function $\chi_j=\chi_{\Si\cap\Si_j}$ and the scalar
$$\begin{array}{rl}
m_j & = \di \frac{1}{|\Si\cap\Si_j|}\int_{\Si\cap\Si_j}
        (\varphi_jg_0+u'\cdot\na'\varphi_j+\be\vp_j u_n)\,dx'\\
    & = \di \frac{1}{|\Si\cap\Si_j|}\int_{\Si\cap\Si_j}(i\xi u_n-g_1)\varphi_j\,dx',
\end{array}$$
we split $g_j$ into the form
$$ g_j=g_{j0}+g_{j1}:=(\varphi_jg_0+u'\cdot \na'\varphi_j+\be\vp_j u_n-m_j\chi_j)+
   (\varphi_jg_1+m_j\chi_j).$$
Concerning $g_{j1}$ we get
$$ \begin{array}{l}
\|g_{j1}\|_{r,\om;\Si_j} \leq \|g_1\|_{r,\om;\Si} + |m_j|\om(\Si\cap\Si_j)^{1/r}\ek
\hspace{1cm} \leq  \|g_1\|_{r,\om;\Si} +
   \di\frac{\om(\Si\cap\Si_j)^{1/r} \cdot \om'(\Si\cap\Si_j)^{1/{r'}}}{|\Si\cap\Si_j|}
   \big(c\|\xi u_n\|_{(W^{1,r'}_{\om'})^*}+\|g_1\|_{r,\om;\Si}\big)
\end{array} $$
where $c>0$ depends on the choice of the cut-off functions $\varphi_j$.
Since we chose the balls
$B_j$ for $j\geq 1$ small enough, for each $j\geq 0$ there is a cube $Q_j$
with $\Si\cap\Si_j\subset Q_j$ and $|Q_j|<c(n)|\Si\cap\Si_j|$
where the constant $c(n)>0$ is independent of $j$. Hence
\begin{equation}
\label{E4.5}
\begin{array}{rl}
\|g_{j1}\|_{r,\om;\Si_j} & \leq  \|g_1\|_{r,\om,\Si} +
   \frac{c(n) \om(Q_j)^{1/r}\cdot\,\om'(Q_j)^{1/{r'}}}
    {|Q_j|} \big(c\|\xi u_n\|_{(W^{1,r'}_{\om'})^*} + \|g_1\|_{r,\om,\Si}\big)\ek
   & \leq c(1+\car^{1/r})\big(\|\xi u_n\|_{(W^{1,r'}_{\om'})^*}+
     \|g_1\|_{r,\om;\Si}\big)
\end{array}
\end{equation}
for $j\geq 0$.
Furthermore, for every test function $\Psi\in C^{\infty}_0(\bar{\Si}_j)$ let
$$ \tilde{\Psi}=\Psi-\frac{1}{|\Si\cap\Si_j|}\int_{\Si\cap\Si_j}\Psi dx'.$$
By the definition of $m_j\chi_j$ we have $\int_{\Si_j}g_{j0}\,dx'=0$;
hence by Poincar\'e's inequality (see Proposition \ref{P2.7})
$$ \begin{array}{l}
\displaystyle \Big|\int_{\Si_j}g_{j0}\Psi \,dx'\Big|
 = \displaystyle\Big|\int_{\Si} \big(g_0(\varphi_j\tilde{\Psi})  +
     u'\cdot(\na'\varphi_j)\tilde{\Psi}
    + \be u_n\varphi_j\tilde{\Psi}\big)\,dx'\Big|\\[2ex]
 \leq \|g_0\|_{-1,r,\om}\|\varphi_j\tilde{\Psi}\|_{1,r',\om'} +
        \|u'\|_{(W^{1,r'}_{\om'})^*}\|(\na'\varphi_j)\tilde{\Psi}\|_{1,r',\om'}
        +\|\be u_n\|_{(W^{1,r'}_{\om'})^*} \|\vp_j \tilde{\Psi}\|_{1,r,\om'}\\[2ex]
 \leq c(\|g_0\|_{-1,r,\om}+\|u\|_{(W^{1,r'}_{\om'})^*})\|\na'\Psi\|_{r',\om';\Si_j},
\end{array}$$
where $c=c(\be)>0$ is $A_r$-consistent.  Thus
\begin{equation}
\label{E4.6}
\|g_{j0}\|_{-1,r,\om;\Si_j}\leq c\big(\|g_0\|_{-1,r,\om} +
\|u\|_{(W^{1,r'}_{\om'})^*}\big)\quad\text{for  } j\geq 0.
\end{equation}
Summarizing \eq{4.5} and \eq{4.6}, we get for $j\geq 0$
$$\|g_j;\hw^{-1,r}_\om(\Si_j)+L^r_{\om,1/\xi}(\Si_j)\|
\leq c\big(\|u\|_{(W^{1,r'}_{\om'})^*}+\|g; L^r_{0,\om}+ L^r_{\om,1/\xi}\|\big)
$$
with an $A_r$-consistent constant $c=c(r, \car)>0$. In view of $\xi^2>\be^2$ we see that
\begin{equation}
\label{E4.7}
\|g_j;\hw^{-1,r}_\om(\Si_j)+L^r_{\om,1/\xi}(\Si_j)\|
\leq c\big(\|u\|_{(W^{1,r'}_{\om'})^*}+\|g; L^r_{0,\om}+ L^r_{\om,1/\eta}\| \big)
\end{equation}
with an $A_r$-consistent $c=c(r, \car)>0$.

To complete the proof, apply Theorem \ref{T3.1} (i) to \eq{4.2}, \eq{4.3} when $j=0$.
Further use Theorem \ref{T3.1} (ii)
 in \eq{4.2}, \eq{4.3} for $j\geq 1$, but with $\la$ replaced by
$\la +M$ with $M=\la_0+\alpha_0$, where $\la_0=\la_0(\ve,r,\car)$
is the $A_r$-consistent constant indicated in Theorem \ref{T3.1} (ii).
This shift in $\la$ implies that $f_j$ has to be replaced by
$f_j+M\varphi_ju$ and that \eq{3.31} will be used with $\la$ replaced by $\la +M$.
Summarizing \eq{3.9}, \eq{3.31} as well as \eq{4.4}, \eq{4.7} and summing over all $j$
we arrive at \eq{4.1} with the additional terms
$$ I = \|Mu\|_{r,\om} + \|Mu\|_{(W^{1,r'}_{\om'})^*} +
   \|Mg; L^r_{0,\om}+ L^r_{\om,1/\eta}\|  $$
on the right-hand side of the inequality.
Note that $M=M(\ve,r,\car)$ is $A_r$-consistent, $|\eta|\leq \max\{\sqrt{2}|\xi|,\sqrt{2}\beta\}$
and that $g=\div'u'+i\eta u_n$ defines a natural splitting of
$g\in L^r_{0,\om}(\Si)+L^r_\om(\Si)$. Hence Poincar\'{e}'s inequality yields
$$ \begin{array}{rl}
I&\leq M\big(\|u\|_{r,\om;\Si}+\|\div'u'\|_{-1,r,\om}+\|u_n\|_{r,\om;\Si}\big)\ek
&\leq c_1\|u\|_{r,\om;\Si}\leq c_2\|\na' u\|_{r,\om;\Si}
\end{array}$$
with $A_r$-consistent constants $c_i=c_i(\ve,r,\Si,\car)>0,\, i=1,2$.

Thus \eq{4.1} is  proved.
\par
\medskip
Next, consider the case $\xi^2\leq \be^2$.
Since $S(u,p)=(f,-g)$, we have
\begin{equation}
\label{E3.3}
\begin{array}{rl}
(\la-\Da')u'+\na' p = f'-\eta^2u', & \div'u'=g-i\eta u_n \quad\text{in }\Si,\ek
& u'|_{\pa\Si}=0,
\end{array}
\end{equation}
and
\begin{equation}
\label{E3.4}
(\la-\Da')u_n=f_n-\eta^2 u_n-i\eta p,\quad \text{in }\Si, \quad  u_n|_{\pa\Si}=0.
\end{equation}
Now apply \cite[Theorem 3.3]{Fr01} to \eq{3.3}.   Then,  in view of
$|\eta|\leq \sqrt{2}\be$ and  Poincar\'e's inequality, for all
$\la\in -\al+S_\ve$, $\al\in (0,\al_0-\be^2)$ we have
$$ \begin{array}{l}
\|(\la+\al)u', \na'^2u', \na'p\|_{r,\om;\Si}\ek \leq
c\big(\|f,\eta^2u\|_{r,\om;\Si}+|\la|\|g-i\eta
u_n\|_{\hat{W}^{-1,r}_\om(\Si)} +\|g-i\eta
u_n\|_{{W}^{1,r}_\om(\Si)}\big)\ek \leq c\big(\|f, \na'u,
p\|_{r,\om;\Si}+\|g\|_{{W}^{1,r}_\om(\Si)} + |\la|\|g-i\eta
u_n\|_{\hat{W}^{-1,r}_\om(\Si)} \big)
\end{array}$$
with $A_r$-consistent constants $c=c(r,\ve,\al, \be, \Si, {\cal
A}_r(\om))$.

In order to control $\|g-i\eta u_n\|_{\hat{W}^{-1,r}_\om(\Si)}$, let
us split $g$ as $g=g_0+g_1$, $g_0\in L^r_{0,\om}(\Si)$, $g_1\in
L^r_{\om,1/\eta}(\Si)$. Since $g_1-i\eta u_n$ has mean value zero in
$\Si$, we get for all $\psi\in C^\infty(\bar\Si)$ and
$\bar\psi=\psi-\frac{1}{|\Si|}\int_{\Si} \psi\,dx'$ by Poincar\'e's
inequality that
$$ \begin{array}{l}
|\lan g_1-i\eta u_n, \psi\ran| = |\lan g_1-i\eta u_n, \bar\psi\ran|\ek
\;\;\; \leq |\eta|\big(\|g_1/\eta\|_{r,\om}\|\|\bar\psi\|_{r',\om'}
  + \|u_n\|_{(W^{1,r'}_{\om'}(\Si))^*}\|\bar\psi\|_{W^{1,r'}_{\om'}(\Si)}\big) \ek
\;\;\; \leq c(r,\Si) \big(\|g_1/\eta\|_{r,\om}
+ \|u_n\|_{(W^{1,r'}_{\om'}(\Si))^*}\big)  \|\na'\psi\|_{r',\om';\Si} .
\end{array}
$$
Therefore,
$$
\|g-i\eta u_n\|_{\hat{W}^{-1,r}_\om}\leq \|g_0\|_{\hat{W}^{-1,r}_\om}
+ c\big(\|g_1/\eta\|_{r,\om}\|+\|u_n\|_{(W^{1,r'}_{\om'})^*}\big). $$
Thus, for all $\la\in -\al+S_\ve$, $\al\in (0,\al_0-\be^2)$ we have
\begin{equation}
\label{E3.5}
\begin{array}{l}
\|(\la+\al)u', \na'^2u', \na'p\|_{r,\om}\ek
\leq c \big( \|f, \na'u, p\|_{r,\om}+\|g\|_{{W}^{1,r}_\om}
+ \|\la u\|_{(W^{1,r'}_{\om'})^*} + \|\la g: L^r_{0,\om}+L^r_{\om,1/\eta}\| \big)
\end{array}
\end{equation}
with
$A_r$-consistent constant $c=c(r,\ve,\al, \be, \Si, {\cal A}_r(\Om))$.

On the other hand, applying well-known results
for the Laplace resolvent equations (cf. \cite{Fr01}) to \eq{3.4}, we get that
\begin{equation}
\label{E3.6}
\|(\la+\al)u_n, \na'^2u_n\|_{r,\om;\Si}
\leq c(\|f_n, u, p\|_{r,\om;\Si}
\end{equation}
 with $c=c(r,\ve,\al, \be, \Si, {\cal A}_r(\Om))$.
Thus, from \eq{3.5} and \eq{3.6} the assertion of the lemma for the case $\xi^2\leq \be^2$ is proved.

The proof of the lemma is complete. \qed

\vspace {0.1cm}
%
%
\begin{lem}
\label{L3.7}
Under the assumptions on $r,\om,\al,\ve$ and $\be,\xi,\la$ as in Lemma \ref{L3.4}
 but with $\al_0$ replaced by $\overline \al=\min\{\al_0,\al_1\}$
there is an $A_r$-consistent constant
$c>0$ such that for $(u,p)\in {\cal D}(S)$, $S= S_{r,\la,\eta}^\om$,
and $(f,-g)=S(u,p)$ the estimate
\begin{equation}
\label{E4.8}
\begin{array}{l}
\|\mu^2_+u,\mu_+\na'u,\na'^2u,\na'p,\eta p\|_{r,\om}\\[1ex]
\hspace{15mm} \leq c\big(\|f,\na'g,g,\xi g\|_{r,\om}+
(|\la|+1)\|g; L^r_{0,\om}+ L^r_{\om,1/\eta}\| \big)
\end{array}
\end{equation}
holds; here $\mu_+=|\la+\alpha+\xi^2|^{1/2}$.
\end{lem}
{\bf Proof:}
Assume that this lemma is wrong.  Then there  is a constant $c_0>0$, a sequence
$\{\om_j\}^\infty_{j=1}\subset A_r$ with ${\cal A}_r(\om_j)\leq c_0$ for all $j$,
sequences $ \{\la_j\}^\infty_{j=1}\subset -\alpha+S_\ve,$ $\{\xi_j\}^\infty_{j=1}\subset \R^*$
and $(u_j,p_j)\in {\cal D}(S^{\om_j}_{r,\la_j,\xi_j})$ for all $j\in\N$ such that
\begin{equation}
\label{E4.9}
\begin{array}{l}
\|(\la_j+\alpha+\xi^2_j)u_j,(\la_j+\alpha+\xi^2_j)^{1/2}\na'u_j,
   \na'^2u_j,\na'p_j,\eta_jp_j\|_{r,\om_j}\ek
\hspace{1cm}\geq j\ \big(\|f_j,\na'g_j,g_j,\xi_jg_j\|_{r,\om_j} +
   (|\la_j|+1)\|g_j; L^r_{0,\om_j}+ L^r_{\om_j,1/\eta_j}\| \big)
\end{array}
\end{equation}
where $\eta_j=\xi_j+i\be$, $(f_j,-g_j)=S^{\om_j}_{r,\la_j,\eta_j}(u_j,p_j)$.

Fix an arbitrary cube $Q$ containing $\Si$. We may assume without
loss of generality  that ${\cal A}_r(\om_j)\leq c_0,\; \om_j(Q)=1$
for all $j\in \N,$ by using the $A_r$-weight
$\tilde{\om}_j:=\om_j(Q)^{-1}\om_j$ instead of $\om_j$ if necessary.
Hence also ${\cal A}_{r'}(\om_j')\leq c_0^{r'/r},\; \om_j'(Q) \leq
c_0^{r'/r} |Q|^{r'}.$ Therefore, by a minor modification of
Proposition \ref{P2.5}, there exist numbers $s, s_1\in (1,\infty)$
such that $L^r_{\om_j}(\Si)\hookrightarrow L^s(\Si)$ and
$L^{s_1}(\Si) \hookrightarrow L^{r'}_{{\om}'_j}(\Si)$ with embedding
constants independent of $j\in\N$.

Furthermore, we may assume without loss of generality that
\begin{equation}
\label{E4.12}
\|(\la_j+\alpha+\xi^2_j)u_j,(\la_j+\alpha+\xi^2_j)^{1/2}
   \na'u_j,\na'^2u_j,\na'p_j,\eta_jp_j\|_{r,\om_j}=1
\end{equation}
and consequently that
\begin{equation}
\label{E4.13}
\|f_j,\na'g_j,g_j,\xi_jg_j\|_{r,\om_j} +
(|\la_j|+1)\|g_j; L^r_{0,\om_j}+ L^r_{\om_j,1/\eta_j}\| \rightarrow 0
    \quad \text{as} \quad j\rightarrow \infty.
\end{equation}
By the above embeddings we conclude from \eq{4.12} that
\begin{equation}
\label{E4.14}
\|(\la_j+\alpha+\xi^2_j)u_j,(\la_j+\alpha+\xi^2_j)^{1/2}
   \na'u_j,\na'^2u_j,\na'p_j,\eta_jp_j\|_s \leq K,
\end{equation}
with some $K>0$ for all $j\in\N$ and from \eq{4.13}
\begin{equation}
\label{E4.15}
\|f_j,\na'g_j,g_j,\eta_jg_j\|_s\rightarrow 0 \quad\text{as}
   \quad j\rightarrow \infty.
\end{equation}
Without loss of generality let us suppose that as $j\rightarrow \infty$,
$$\begin{array}{l}
\la_j\rightarrow \la\in -\alpha+\bar{S}_\ve \quad \text{or}
   \quad|\la_j|\rightarrow \infty\\
\xi_j\rightarrow 0  \quad \text{or}\quad \xi_j\rightarrow \xi \neq 0
   \quad \text{or}\quad|\xi_j|\rightarrow \infty.
\end{array}$$
Thus we have to consider six possibilities, each of them leading to a contradiction
as in the proof of \cite[Lemma 4.3]{FR05-3}.

The first three cases are $\la_j\rightarrow \la \in -\alpha+\bar{S}_\ve,\quad
\xi_j\rightarrow \xi \in\overline\R$, cf. Case (i), (ii) and (iii) in \cite[Lemma 4.3]{FR05-3};
these cases are analyzed in a completely analogous way where even
 the case $\xi=0$ poses no difficulties since $\eta=\xi+i\be \neq 0$.\\[1ex]


Let us consider more carefully the
Case (iv) $|\la_j|\rightarrow \infty,$ $\xi_j\rightarrow \xi\in\R$:\\
We follow Case (iv) in \cite[Lemma 4.3]{FR05-3} and argue as follows:
By \eq{4.12}
\begin{equation}
\label{E4.22}
\|\na'u_j,\xi_ju_j\|_{r,\om_j}\rightarrow 0\quad \text{as}\quad j\rightarrow \infty.
\end{equation}
Further, \eq{4.14} yields the convergence
$$\begin{array}{lcl}
u_j\rightarrow 0, \na'u_j\rightarrow 0 & \quad \text{and}\quad
& \na'^2u_j\wrar 0, \la_ju_j\wrar v,\\
 p_j\rightarrow p&\quad\text{and} \quad & \na'p_j\wrar \na'p,
\end{array}$$
in $L^s$, which, together with \eq{4.15}, leads to
\begin{equation}
\label{E4.23}
v'+\na'p=0,\quad v_n+i\eta p=0.
\end{equation}

From \eq{4.13} w find a splitting $g_j = g_{j0}+g_{j1}$, $g_{j0}\in L^r_{0,\om_j}$,
$g_{j1}\in L^r_{\om_j}$ such that
\begin{equation}
\label{E4.17}
\|\la_j g_{j0}\|_{-1,r,\om_j}+\|\la_j g_{j1}/\eta_j\|_{r,\om_j}\ra 0\quad (j\ra\infty)
\end{equation}
and
\begin{equation} \begin{array}{rl}
|\lan\la_jg_j,\varphi\ran|
& = |\lan\la_jg_{j0},\varphi\ran + \lan\la_jg_{j1},\varphi\ran| \nn\ek
& \leq \|\la_jg_{j0}\|_{-1,r,\om_j}\|\na'\varphi\|_{r',\om'_j} +
  \|\la_jg_{j1}\|_{r,\om_j}\|\varphi\|_{r',\om'_j}\label{E4.23n} \ek
& \leq c \big(\|\la_jg_{j0}\|_{-1,r,\om_j} +
  \|\la_jg_{j1}/\eta_j\|_{r,\om_j}\big)\|\varphi\|_{W^{1,s_1}(\Si)}. \nn
\end{array} \end{equation}
Consequently, due to \eq{4.17},
\begin{equation}
\label{E4.24}
\la_jg_j\in(W^{1,s_1}(\Si))^*\quad\text{and}
\quad\| \la_jg_j\|_{(W^{1,s_1}(\Si))^*}\rightarrow 0
\quad \text{as }j\rightarrow \infty.
\end{equation}
Now the divergence equation $\div\hspace{-0.1cm}_{\eta_j}' u_{j} = g_{j}$ implies
that for all $\varphi\in C^\infty(\bar{\Si})$
$$ \begin{array}{rcl}
\lan v',-\na'\varphi\ran + \lan i\eta v_n,\varphi \ran
& = & \lim_{j\rightarrow \infty}\lan\div'\la_j u'_j +
      i\la_j\eta_j u_{jn},\varphi\ran\ek
& = & \lim_{j\rightarrow \infty}\lan\la_jg_j,\varphi\ran=0 ,
\end{array}$$
yielding $\div\hspace{-0.1cm}'v' = -i\eta v_n,\; v'\cdot N|_{\pa\Si}=0.$
Therefore \eq{4.23} leads to the Neumann problem
\begin{equation}
\label{E3.7}
-\Da'p+\eta^2p=0\text{ in }\Si, \quad \di \frac{\pa p}{\pa N}=0
   \text{ on }\pa\Si.
\end{equation}
Here note that $\eta^2=\xi^2-\be^2+2i\xi\be$.
Hence, if $\xi\neq 0$ then $p\equiv 0$ since the eigenvalues of the
Neumann Laplacian in $\Si$ are real; if $\xi=0$, then
$\eta^2=-\be^2$ and hence $p\equiv 0$ due to the condition
$0<\be^2<\bar\al\leq \al_1$. Consequently, $p\equiv 0$ and also $v\equiv 0$.

Now, due to Proposition \ref{P2.6} (2), (3), we get the convergences
$\|\la_j u_j\|_{(W^{1,r'}_{\om_j'})^*} \to 0$ and
$\|p_j\|_{r,\om_j}\rightarrow 0$ as $j\to \infty$,
since $\la_ju_j\wrar 0$ in $L^s$, $p_j\wrar0$ in $W^{1,s}$ and
$\sup_{j\in\N} \|\la_ju_j\|_{r,\om_j}<\infty,$
$\sup_{j\in\N} \|p_j\|_{1,r,\om_j}<\infty$.
Thus \eq{4.1}, \eq{4.12}, \eq{4.13} and \eq{4.22} lead to the contradiction
$1\leq 0$.

The last case (vi) in which $|\la_j|\rightarrow \infty$ and $|\xi_j|\rightarrow \infty$
is analyzed as Case (vi) in \cite[Lemma 4.3]{FR05-3} with only minor modifications.

Now the proof of this lemma is complete. \hfill \qed

%
%
\begin{theo}
\label{T3.8}
Let $1<r<\infty,\, \om\in A_r$ and
$\xi\in\R^*$, $\be \in (0, \sqrt{\bar\al})$, $\al\in (0,\bar\al-\be^2)$,
 $\ve\in \big(0,\arctan(\frac1\be \sqrt{\bar\al-\be^2-\al} )\big)$.
Then for every $\la\in-\alpha+S_\ve$, $\xi\in \R^*$ and
$f\in L^r_\om(\Si),\, g\in W^{1,r}_\om(\Si)$ the parametrized resolvent problem
$(R_{\la,\xi,\be})$ has a unique solution
$(u,p)\in \big(W^{2,r}_\om(\Si)\cap W^{1,r}_{0,\om}(\Si)\big)\ti W^{1,r}_\om(\Si)$.
Moreover, this solution satisfies the estimate \eq{4.8}
with an $A_r$-consistent constant $c=c(\alpha,\be,\ve,r,\Si,\car)>0$.
\end{theo}
{\bf Proof:} The existence is obvious since, for every
$\la\in-\alpha+S_\ve,\xi\in \R^*$ and $\om\in A_r(\R^{n-1})$, the range
${\cal R}(S^\om_{r,\la,\xi})$ is closed and dense in
$L^r_\om(\Si)\ti W^{1,r}_\om(\Si)$ by Lemma \ref{L3.7} and by Lemma \ref{L3.4}, respectively.
Here note that for fixed $\la\in \C,\,\xi\in\R^*$ the norm
$\|\na'g,g,\xi g\|_{r,\om}+ (1+|\la|) \|g; L^r_{0,\om}+ L^r_{\om,{1/\xi}}\| $
is equivalent to the norm of $W^{1,r}_\om(\Si)$.
The uniqueness of solutions is obvious from Lemma \ref{L3.4}. \vspace{0.4cm}\hfill\qed

Now, for fixed $\om\in A_r, 1<r<\infty$, define the
operator-valued functions
$$\begin{array}{l}
a: \R^*\rightarrow {\cal L}(L^r_\om(\Si);W^{2,r}_{\om}(\Si)
     \cap W^{1,r}_{0,\om}(\Si)),\\[1ex]
b: \R^*\rightarrow  {\cal L}(L^r_\om(\Si);W^{1,r}_\om(\Si))
\end{array}$$
by
\begin{equation}
\label{E4.26}
 a(\xi)f:=u(\xi),\quad b(\xi)f:=p(\xi),
\end{equation}
where $(u(\xi), p(\xi))$ is the solution to $(R_{\la,\xi,\be})$
corresponding to $f\in L^r_\om(\Si)$ and $g=0$.
%

\begin{coro}
\label{C3.9} Assume the same for $\alpha,\be,\xi,\ve,\la$ as in
Theorem \ref{T3.8}. Then, the operator-valued functions $a, b$
defined by \eq{4.26} are Fr\'{e}chet differentiable in $\xi\in\R^*$.
Furthermore, their derivatives $w = \frac{d}{d\xi}a(\xi)f,\; q =
\frac{d}{d\xi}b(\xi)f$ for fixed $f\in L^r_\om(\Si)$ satisfy the
estimate
\begin{equation}
\label{E4.28} \|(\la+\alpha)\xi w,\xi\na'^2w, \xi^3w,
\xi\na'q,\xi\eta q\|_{r,\om} \leq c\|f\|_{r,\om}
\end{equation}
with an $A_r$-consistent constant $c=c(\alpha, \be, r,\ve,\Si,\car)$
independent of $\la\in -\alpha+S_\ve$ and $\xi\in\R^*$.
\end{coro}
{\bf Proof:} Since $\xi$ enters in $(R_{\la,\xi})$ in a polynomial
way, it is easy to prove that $a(\xi),b(\xi)$ are Fr\'{e}chet
differentiable and their derivatives $w,q$ solve the system
\begin{equation}
\label{E4.30}
\begin{array}{rcl}
(\la+\eta^2-\Da')w'+\na'q & = & -2\eta u'\\[1ex]
(\la+\eta^2-\Da')w_{n}+i\eta q & = & -2\eta u_{n}-ip\\[1ex]
\div'w'+i\eta w_{n} & = & -iu_{n},
\end{array}
\end{equation}
where $(u,p)$ is the solution to $(R_{\la,\xi,\be})$ for $f\in
L^r_\om(\Si),\, g=0$.

We get from \eq{4.30} and Theorem \ref{T3.8} that
\begin{equation}
\label{E4.31}
\begin{array}{l}
\|(\la+\alpha)\xi w,\xi\na'^2w, \xi^3w, \xi\na'q,\xi\eta q\|_{r,\om}\\[1ex]
\qquad \leq c\big(\|\xi\eta u,\xi p, \xi\na'u_{n}, \xi^2 u_{n}\|_{r,\om}
       + (|\la|+1)\|i\xi u_{n}; L^r_{0,\om}+ L^r_{\om,1/\eta}\| \big)\\[1ex]
\qquad \leq c\big(\|\xi^2 u,\xi p, \xi\na' u\|_{r,\om}
       + (|\la|+1)\|u\|_{r,\om}\big)\ek
\qquad \leq c \big\|u, (\la+\alpha+\xi^2)u,
   \sqrt{\la+\alpha+\xi^2}\,\na' u, \xi p\big\|_{r,\om}\ek
\qquad \leq c \big\|(\la+\alpha+\xi^2)u,
   \sqrt{\la+\alpha+\xi^2}\,\na' u, \na'^2 u,\xi p\big\|_{r,\om},
\end{array}
\end{equation}
with an $A_r$-consistent constant $c=c(\alpha,r,\ve,\Si,\car)$; here
we used the fact that $\xi^2+ |\la+\alpha|\leq c(\ve)
|\la+\alpha+\xi^2|$  for all $\la\in -\alpha+S_\ve, \xi\in \R$, then $|\xi|\leq|\eta|\leq |\xi|+\sqrt{\bar{\al}}$ and
$\|u\|_{r,\om}\leq c(\car)\|\na'^2u\|_{r,\om}$, see \cite[Corollary 2.2]{Fr01}. Thus Theorem \ref{T3.8} and \eq{4.31} yield
\eq{4.28}. \hfill \qed

%
%
\section{Proof of the Main Results}
The proof of Theorem 2.1 is based on the theory of operator-valued
Fourier multipliers.
The classical  H\"{o}rmander-Michlin theorem for scalar-valued multipliers
for $L^q(\R^k),\, q\in(1,\infty),\, k\in\N,$ extends to an operator-valued
version for Bochner spaces $L^q(\R^k;X)$ provided that $X$ is a {\it UMD space}
and that the boundedness condition for the derivatives of the multipliers
is strengthened to {\it ${\cal R}$-boundedness}.

Recall that a Banach space $X$ is called a {\it UMD} space if the Hilbert transform
on the Schwartz space of all rapidly
decreasing $X$-valued functions extends to a bounded linear operator
in $L^q(\R;X)$ for some $q \in(1,\infty)$ (and then even for all $q\in(1,\infty)$,
see e.g. \cite[Theorem 1.3]{RRT86}). We note that
weighted Lebesgue spaces $L^r_\om(\Si)$, $1<r<\infty,\,\om\in A_r,$
are {\it UMD} spaces.
%
%
%
%
%
\begin{tdefi}
\label{D4.2}
Let $X,Y$ be Banach spaces. An operator family ${\cal T}\subset{\cal L}(X;Y)$
is called $\cal R$-bounded if there is a constant $c>0$ such that for all
$T_1,\ldots,T_N\in{\cal T},$ $x_1,\ldots, x_N\in X$ and $N\in\N$
\begin{equation}
\label{E5.1}
\Big\|\sum_{j=1}^{N}\ve_j(s)T_jx_j\Big\|_{L^q(0,1;Y)}
\leq c\, \Big\|\sum_{j=1}^{N}\ve_j(s)x_j\Big\|_{L^q(0,1;X)}
\end{equation}
for some $q\in [1,\infty)$, where $(\ve_j)$ is any sequence of independent,
symmetric $\{-1,1\}$-valued random variables on $[0,1]$.
The smallest constant $c$ for which {\rm \eq{5.1}} holds is denoted by
$R_q({\cal T})$, the $\cal R$-bound of ${\cal T}$.
\end{tdefi}

%

%

We recall an operator-valued Fourier multiplier theorem in Banach spaces.

%
\begin{theo} {\em (\cite[Theorem 3.19]{DHP03}, \cite[Theorem 3.4]{We01})}
\label{T4.6}
Let $X$ and $Y$ be UMD spaces and $1<q<\infty$.
Let $M: \R^*\rightarrow {\cal L}(X,Y)$ be a differentiable function
such that
$$ {\cal R}_q \big(\{M(t),\,tM'(t): \,\,t\in\R^*\}\big) \leq A.$$
Then the operator
$$ Tf = \big(M(\cdot)\hat{f}(\cdot)\big)^\vee,\quad f \in C_0^\infty(\R^*;X),$$
 extends to a bounded operator $T: L^q(\R;X)\rightarrow L^q(\R;Y)$
with operator norm $\|T\|_{{\cal L}(L^q(\R;X);L^q(\R;Y))}\leq CA$
where $C>0$ depends only on $q,X$ and $Y$.
\end{theo}

\begin{rem}
\label{R4.7} For $X=L^r_\om(\Si),\, 1<r<\infty,\,\om\in A_r,$ the
constant $C$ in Theorem \ref{T4.6} is independent of the weight
$\om$, see \cite[Remark 5.7]{FR05-3}.
\end{rem}
       %
%
Now we are in a position to prove Theorem 2.1.\vspace{0.3cm}\\
%
{\bf Proof of Theorem 2.1:}
 Let $f(x',x_n):= e^{\be x_n}F(x',x_n)$ for
 $(x',x_n)\in\Si\ti\R$
 and let us define $u,p$ in the cylinder
$\Om=\Si\ti \R$ by
$$ u(x)={\cal F}^{-1}(a \hat{f})(x),
   \quad  p(x)={\cal F}^{-1}(b \hat{f})(x),$$
where $a, b$ are the operator-valued multiplier functions defined in
\eq{4.26}.

For $\xi\in \R^*$ define $m_\la(\xi):L^r_\om(\Si)\rightarrow
L^r_\om(\Si)$ by
$$ m_\la(\xi)f := \big((\la+\alpha)a(\xi)\hat{f}, \xi\na'a(\xi)\hat{f},
   {\na'}^2 a(\xi)\hat{f}, \xi ^2a(\xi)\hat{f},
   \na'b(\xi)\hat{f},(\xi+i\be) b(\xi)\hat{f}\big).$$
Theorem \ref{T3.8} and Corollary \ref{C3.9} yield the estimate
$$ \sup_{\xi\in \R^*}\|m_\la(\xi), \xi m'_\la(\xi)
   \|_{{\cal L}(L^r_\om(\Si))}<c(q,r,\al,\be,\ve,\Si, \car) $$
for any Muckenhoupt weight $\om\in A_r(\R^{n-1})$.
Therefore, by an {\it extrapolation theorem} (cf. \cite[Theorem 5.8]{FR05-3})
the operator family
$\{m_\la(\xi), \xi m'_\la(\xi):\xi\in \R^*\}$ is $\cal R$-bounded in
${\cal L}(L^r_\om(\Si))$; to be more precise,
$$ {\cal R}_q\big(\{m_\la(\xi), \xi m'_\la(\xi):\xi\in \R^*\}\big)
   \leq c(q,r,\alpha,\be, \ve,\Si,\car) < \infty.$$
Hence Theorem \ref{T4.6} and Remark \ref{R4.7} imply that
$$ \|(m_\la\hat{f})^\vee\|_{L^q(L^r_\om)} \leq C\|f\|_{L^q(L^r_\om)}$$
with an $A_r$-consistent constant $C=C(q,r,\alpha,\be,
\ve,\Si,\car)>0$ independent of the resolvent parameter $\la\in
-\alpha+S_\ve$. Therefore, by the definition of the multiplier
$m_\la(\xi)$, we have $(\la+\alpha)u,\na^2u, \na' p, (\pa_n-\be)p\in
L^q(L^r_\om)$ and
\begin{equation}
\label{E5.5}
 \|(\la+\alpha)u,\na^2 u,\na' p,
(\pa_n-\be)p\|_{L^q(L^r_\om)}
   \leq \|(m_\la\hat{f})^\vee\|_{L^q(L^r_\om)}\leq C\|f\|_{L^q(L^r_\om)},
   \end{equation}
which, in particular, implies by Poincar\'e's inequality
 \begin{equation}
\label{E5.6} u \in W^{2;q,r}_\om(\Om)\cap
W^{1;q,r}_{0,\om}(\Om),\quad \|u\|_{W^{2;q,r}_\om(\Om)} \leq
C\|f\|_{L^q(L^r_\om)}.
\end{equation}
 Note that $(u, p)$ is the solution to the system
$$ (\la-\Da)u - (\be^2-2\be \pa_n)u + \big(\na',\pa_n-\be\big)^\bot p = f,
   \quad \div u-\be u_n=0,$$
which, after being multiplied by $e^{-\be x_n}$, implies that
$(U, P):=(e^{-\be x_n}u, e^{-\be x_n} p)$ solves $(R_\la)$ with
$F=e^{-\be x_n}f$, $G=0$ and satisfies
$$ (\la+\al)U, \na^2U,\na P\in L^q_\be(L^r_\om)$$
as well as the estimate \eq{2.1} in view of \eq{5.5} and \eq{5.6}.

Thus the existence of a solution satisfying \eq{2.1} is proved.

For the proof of uniqueness let $(U,P)$ be a solution of the homogeneous problem $(R_\la )$
such that $(\la+\al)U, \na^2U,\na P\in L^q_\be(L^r_\om)$.
Moreover, let $u=e^{\be x_n}U, p=e^{\be x_n}P$. Then, for a.a.
$\xi \in \R$, $(\hat{u}(\xi),\hat{p}(\xi))\in
\big(W^{2,r}_\om(\Si)\cap W^{1,r}_{0,\om}(\Si)\big)\ti
W^{1,r}_\om(\Si)$ is the solution to $(R_{\la,\xi,\be})$ with
$f=g=0$, and hence $(\hat{u}(\xi),\hat{p}(\xi))=0$ by \eq{4.8}.
Thus we have $U=0, \na P=0$, and the proof of Theorem 2.1
is complete. \hfill \qed\\[2ex]
%
%
 {\bf Proof of Corollary 2.2:} Defining the Stokes operator $A=A_{q,r;\be,\om}$
 by \eq{2.2}, due to the Helmholtz decomposition of the space
 $L^q_\be(L^r_\om)$ on the cylinder $\Om$, see \cite{Fa03},
 we get that for $F\in L^q_\be(L^r_\om)_\si$ the solvability of the equation
\begin{equation}
\label{E5.7}
(\la+A)U=F \quad \text{in}\quad L^q_\be(L^r_\om)_\si
\end{equation}
is equivalent to the solvability of $(R_\la)$ with right-hand side
$G\equiv 0$. By virtue of Theorem 2.1 for every
$\la\in -\alpha+S_\ve$ there exists a unique solution
$U=(\la+A)^{-1}F\in \mathcal D(A)$ to \eq{5.7} satisfying the estimate
$$ \|(\la+\alpha)U\|_{L^q_\be(L^r_\om)_\si}
=\|(\la+\alpha)u\|_{L^q(L^r_\om)}
\leq C\|f\|_{L^q(L^r_\om)}=C\|F\|_{L^q_\be(L^r_\om)_\si}$$
with $C=C(q,r,\alpha,\be,\ve,\Si,\car)$ independent of $\la$,
where $u=e^{\be x_n}U$, $f=e^{\be x_n}F$.
Hence \eq{2.3} is proved.
Then \eq{2.4} is a direct consequence of \eq{2.3} using semigroup theory.
\qed
\par\bigskip

\noindent {\bf Proof of Theorem 2.3:}
Let us show that the operator family
$${\cal T}=\{\la(\la+A_{q,r;\be,\om})^{-1}: \;\la\in \mathrm{i}\R\}$$
is $\cal R$-bounded in ${\cal L}(L^q_\be(L^r_\om)_\si)$. By the way,
since $L^q_\be(L^r_\om)_\si$ is isomorphic to a closed subspace $X$
of $L^q(L^r_\om)$ with isomorphism $I_\be F:= e^{\be x_n}F$, it is
enough to show ${\cal R}$-boundedness of the family
$${\cal \tilde{T}}=\{I_\be\la(\la+A_{q,r;\be,\om})^{-1}I_\be^{-1}: \;\la\in \mathrm{i}\R\}
\subset {\cal L}(X).$$

For $\xi\in\R^*$ and $\la\in S_\ve$,
 let $m_\la(\xi):=\la a(\xi)$ where $a(\xi)$
is the solution operator for $(R_{\la,\xi,\be})$ with $g=0$ defined by \eq{4.26}.
Then, we have
 $$I_\be\la(\la+A_{q,r;\be,\om})^{-1}I_\be^{-1} f= \la I_\be U=(m_\la(\xi)\hat{f})^\vee,
                  \quad \forall f\in X,$$
where $U$ is the solution to $(R_\la)$ with $F=I_\be^{-1}f$, $G=0$.
Hence,  ${\cal R}$-boundedness of ${\cal \tilde{T}}$ in ${\cal
L}(X)$ is proved if there is a constant $C>0$ such that
\begin{equation}
\label{E5.9}
\Big\|\sum_{i=1}^N\ve_i(m_{\la_i}\hat{f}_i)^\vee
      \Big\|_{L^q(0,1;L^q(L^r_\om))}
\leq C \Big\|\sum_{i=1}^N\ve_i f_i\Big\|_{L^q(0,1;L^q(L^r_\om))}
\end{equation}
for any independent, symmetric and $\{-1,1\}$-valued random
variables $(\ve_i(s))$ defined on $(0,1)$, for all $(\la_i)\subset
\mathrm{i} \R$ and $(f_i)\subset X$. Note that we have ${\cal
R}$-boundedness of the operator family $\{m_\la(\xi),\xi
m'_\la(\xi):\xi\in\R^*\}$ in ${\cal L}(L^r_\om)$ due to Theorem
\ref{T3.8}, Corollary \ref{C3.9} and the extrapolation theorem (cf.
\cite[Theorem 5.8]{FR05-3}). Using this property, \eq{5.9} can be
proved via Schauder decomposition approach exactly in the same way
as the proof of \cite[(5.7), pp. 384-386]{FR05-3} in the proof of
\cite[Theorem 2.3]{FR05-3}; hence we omit it.

Then, by \cite[Corollary 4.4]{We01}, for each $f\in L^p(\R_+;
L^q_\be(L^r_\om)_\si), 1<p<\infty,$ the mild solution $U$ to the
system
\begin{equation}
\label{E5.14}
U_t+A_{q,r;\be, \om}U=F,\quad u(0)=0
\end{equation}
belongs to
$L^p(\R_+; L^q_\be(L^r_\om)_\si)\cap L^p(\R_+; D(A_{q,r;\be,\om}))$
and satisfies the estimate
$$ \| U_t, A_{q,r;\be,\om}U\|_{L^p(\R_+; L^q_\be(L^r_\om)_\si)}
   \leq C\|F\|_{L^p(\R_+; L^q_\be(L^r_\om)_\si)}.$$
Furthermore \eq{2.3} with $\la=0$ implies that also $U$ obeys
this inequality thus proving \eq{2.5b}. The remaining part of the proof is easy;
for \eq{2.5c} we use the Helmholtz projection
in $L^q_\be(L^r_\om)$ (see \cite{Fa03}), and for \eq{2.5d}
we work with the new unknown $V(t)=e^{\al t}U(t)$ leading to a spectral shift by $\al$.

The proof of Theorem 2.3 is complete. \hfill\qed

\bigskip
\noindent
{\bf Proof of Theorem \ref{T2.4}:}
Let $1<q<\infty$ and
$\xi\in\R^*$, $\be \in (0, \sqrt{\al^*})$, $\al^*=\min_{1\leq i\leq
m}{\bar\al_i}$, $\al\in (0,\al^*-\be^2)$,
 $\ve\in \big(0,\arctan(\frac{1}{\be}\sqrt{\al^*-\be^2-\al})\big)$.
Fix $\la\in-\alpha+S_\ve$ and $\xi\in \R^*$. Note that
$\la+A_{q,\tb}$ with $\be_i=0$ for all $i=1,\ldots,m$ is injective
and surjective, see \cite[Theorem 1.2]{FR05-4}.
Hence, given any $F\in  L^q_{\tb,\si}(\Om) \subset L^q_\sigma(\Om)$, for all $\la\in -\al+S_\ve$
there is a unique $(U, \na P)\in D(A_q)\ti L^q(\Om)$ such that
\begin{equation}
\label{E4.40}
 \begin{array}{rcll}
\la U- \Da U +  \na P & =& F & \mbox{ in } \Om,\ek
\div U & = & 0 & \mbox{ in }\Om,\ek
U & = & 0 & \mbox{ on } \pa \Om.
\end{array}
\end{equation}

Without loss of generality we may assume that there exist cut-off functions
$\{\varphi_i\}_{i=0}^{m}$ such that
\begin{equation}
\label{E4.15n}
\begin{array}{l}
    \sum_{i=0}^{m}\varphi_i(x)=1, \quad 0\leq \varphi_i(x)\leq 1\quad
    \text{for }x\in\Om,\ek
\varphi_i \in C^{\infty}(\bar{\Om}_i), \quad
   \text{dist}\,(\text{supp}\,\varphi_i,\, \pa\Om_i\cap \Om)\geq \delta>0,
   \,\,i=0,\ldots, m.
\end{array}
\end{equation}
In the following, for $i=1,\ldots, m$ let $\widetilde{\Om}_i$ be
the infinite straight cylinder extending the semi-infinite cylinder $\Om_i$, and
denote the zero extension of $\vp_i v$ to $\widetilde{\Om}_i$ by $\widetilde{\vp_iv}$; furthermore, let $\widetilde{\Om}_0:=\Om_0$ and $\widetilde{\vp_0v} := \vp_0v$.

Define
\begin{equation}\label{E4.7n}
(u^{0},p^{0}): =(\vp_0 U,\vp_0 P),\;
(u^{i}, p^{i}):=(\widetilde{\vp_i U}, \widetilde{\vp_i P})\;\text{ for }i=1,\ldots,m. \end{equation}
Then
$(u^i, p^i)$, $i=0,\ldots, m$, solves on $\widetilde{\Om}_i$ the resolvent problem
\begin{equation} \label{E4.21x}
\begin{array}{rcll}
\la u^i- \Da u^i + \na p^i & =&\tilde{f}^i & \mbox{ in } \widetilde{\Om}_i,\ek
\div u^i & = & \tilde{g}^i & \mbox{ in }\widetilde{\Om}_i,\ek
u^i & = & 0 & \mbox{ on } \pa \widetilde{\Om}_i,
\end{array}
\end{equation}
where
$$ f^i:=\varphi_iF+(\na\varphi_i)P-(\Da\varphi_i)U-2\na\varphi_i\cdot\na U,
   \quad g^i:=\na\varphi_i\cdot U,\quad i=0,\ldots, m.$$
Since  $\,\,\text{supp}\, g^i\subset \Om_0$, $g^i\in
W^{1,q}_0(\Om_0)$ and $\int_{\Om_0}g^i\,dx=0$ for  $i=0,\ldots,m$,
we find due to the well-known theory of the divergence problem
some $w_i\in W^{2,q}_0(\Om_0)$ satisfying $\div w_i=g^i$ in $\Om_0$ and
\begin{equation}
\label{E4.10}
\begin{array}{l}
\|\na^2 w_i\|_{L^q(\Om_0))} \leq c\|\na g^i\|_{L^q(\Om_0)} \leq c\|\na U\|_{L^{q}_0(\Om_0)},\ek
\|w_i\|_{L^q(\Om_0)} \leq c\|g^i\|_{(W^{1,q}(\Om_0))^*} \leq c\|U\|_{(W^{1,q}(\Om_0))^*}
\end{array}
\end{equation}
for  $i=0,\ldots,m$, where $c=c(\Om_0,q)$, cf. \cite[Remarks, p. 274]{FS94MMAS} and
\cite[Chapter III.3]{Ga94-1}.
Although a solution to the problem $\div w_i=g^i$ is not unique, we note that there exists a linear solution operator $g^i\mapsto w^i$, see the explicit construction in \cite[Chapter III, Lemma 3.1]{Ga94-1}.
Then $\tilde{w}_i$, the extension by $0$ of $w_i$ to $\widetilde{\Om}_i$,
$i=1\ldots,m,$ satisfies
\begin{equation}
\label{E4.17n}
\begin{array}{l}
e^{\be_i x^i_n}\na^2 \tilde{w}_i
\in L^q(\tilde\Om_i),\;
 \| e^{\be_i x^i_n}\na^2 \tilde{w}_i\|_{L^q(\tilde\Om_i)}
 \leq c\|\na U\|_{L^{q}(\Om_0)}.
\end{array}
\end{equation}

Now, $v^0:=u^0-w_0$  solves \eq{4.21x} with $\tilde f^0$ replaced by
$f^0-(\la w_0-\Delta w_0)$ and $g^0=0$ so that resolvent estimates for the Stokes problem on bounded domains together with \eq{4.10} yield
\begin{equation}\label{E4.10n}
\|v^0, \la v^0,\na^2 v^0, \na p^0\|_{L^q(\Om_0)} \leq
 c\|F,\na U,P\|_{L^{q}(\Om_0)}+(|\la|+1)\|U\|_{(W^{1,q}(\Om_0))^*}
\end{equation}
with $c$ independent of $\la$.
Moreover, $v^i:=u^i-\tilde{w}_i$, $i=1,\ldots,m$, solve \eq{4.21x}
with $\tilde f^i$ replaced by $\tilde f^i - (\lambda\tilde w_i -\Delta \tilde w_i)$
and $\tilde g^i=0$. Hence by Theorem \ref{T2.1} and \eq{4.17n} we have
\begin{equation}\label{E4.11}
\begin{array}{l}
\|v^i, \la v^i, \na^2 v^i, \na p^i\|_{L^q_{\be_i}(\R; L^q(\Si^i))} \\[2ex]
\quad \leq  c\big(\|F\|_{L^{q}_{\be_i}(\tilde\Om_i)} +\|\na U, P\|_{L^{q}(\Om_0)}
             + (|\la|+1)\|U\|_{(W^{1,q}(\Om_0))^*}\big),
\end{array}
\end{equation}
$i=1,\ldots,m$, with $c$ independent of $\la$.
Due to $U=\sum_{i=0}^m u^i$, $P=\sum_{i=0}^m p^i$ in $\Om$ and the estimates \eq{4.17n}-\eq{4.11}, we get $\na^2 U, \na P\in L^q_{\tb}(\Om)$ and
\begin{equation}\label{E4.34}
\begin{array}{l}
\|U, \la U,\na^2 U, \na P\|_{L^q_{\tb}(\Om)} \\[1ex]
\qquad \leq   c\big(\|F\|_{L^{q}_{\tb}(\Om)}+\|\na U, P\|_{L^{q}(\Om_0)}\big)
              + (|\la|+1)\|U\|_{(W^{1,q}(\Om_0))^*}.
\end{array}
\end{equation}

Now we shall show that \eq{4.34} implies, by a contradiction argument, that
\begin{equation}
\label{E4.35}
\begin{array}{l}
\|U, \la U,\na^2 U, \na P\|_{L^q_{\tb}(\Om)} \leq c\|F\|_{L^{q}_{\tb}(\Om)}
\end{array}
\end{equation}
with $c$ independent of $\la$.

Assume that \eq{4.35} does not hold.
Then there are sequences $\{\la_j\}_{j\in\N}\subset -\alpha+S_\ve,$
$\{(U_j,P_j)\}_{j\in\N}$ such that
\begin{equation}
\label{E4.37}
\|U_j, \la_j U_j,\na^2 U_j,\na P_j\|_{L^q_\tb(\Om)}=1,
    \quad\|F_j\|_{L^q_\tb(\Om)}\ra 0 \quad\text{as  }j\ra\infty,
\end{equation}
where $F_j=\la U_j-\Da U_j+\na P_j$, $\div U_j=0$. Without loss of generality we
may assume the following weak convergence in $L^q_\tb(\Om)$:
\begin{equation}
\label{E4.38}
\la_j U_j\wrar V,\; U_j\wrar U, \; \na^2 u_j\wrar \na^2 U, \;
\na P_j\wrar \na P\quad\text{as  }j\ra\infty
\end{equation}
with some $V\in L^q_\tb(\Om),\, U\in W^{2,q}_\tb(\Om)\cap W^{1,q}_{0,\tb}(\Om) \cap L^{q}_{\tb,\sigma}(\Om)$  and $P\in \hw^{1,q}_\tb(\Om)$.
Moreover, we may assume $\int _{\Om_0} P_j \, dx=0$, $j\in\N$, $\int
_{\Om_0} P \, dx=0$ and either $\la_j\ra \la\in
\{-\alpha+\bar{S}_\ve\}$ or $|\la_j|\to\infty$ for $j\ra\infty$.
\vspace{0.2cm}

\par (i) Let $\la_j\ra \la \in -\alpha+\bar{S}_\ve$.
Then, $V=\la U$ and it follows that  $(U,P)$ solves \eq{4.40} with
$F=0$ yielding $(U,P)=0$. On the other hand, using the compact embeddings
$W^{2,q}(\Om_0)\subset\subset W^{1,q}(\Om_0)\subset\subset L^q(\Om_0)                   \subset\subset (W^{1,q'}(\Om_0))^*$
and Poincar\'e's inequality on $\Om_0$, we have the strong convergence
\begin{equation}
\label{E4.39}
U_j\ra 0 \,\,\,\text{in  } W^{1,q}(\Om_0),\,\,\, P_j\ra 0\,\,\,\text{in  }L^q(\Om_0),\,\,\,
  (|\la_j|+1)U_j \ra 0\,\,\,\text{in  }(W^{1,q'}(\Om_0))^* .
\end{equation}
Thus \eq{4.35} yields the contradiction $1\leq 0$.

\par (ii) Let $|\la_j|\ra \infty$.
Then, we conclude that
$U=0$, and consequently $V+\na P=0$ where $V\in L^q_{\si}(\Om)$.
Note that this is the $L^q$-Helmholtz decomposition of the null vector field on $\Om$.
Therefore, $V=0,\, \na P=0$.
Again we get \eq{4.39} and finally the contradiction $1\leq 0$.

Summarizing we proved the resolvent estimate \eq{4.35}. Hence $A_{q,\tb}$ is the generator of an exponentially decaying analytic semigroup on $L^q_{\tb,\sigma}(\Om)$.
\qed

\par\bigskip
\noindent
{\bf Proof of Theorem \ref{T2.5}:}
Note that $L^p(\R_+; L^q_\tb(\Om))\subset L^p(\R_+; L^q(\Om))$ for
$1<p,q<\infty$. Hence, by maximal $L^p$-regularity of the Stokes
operator in $L^q_\sigma(\Om)$, which follows by \cite[Theorem 1.2]{FR05-4},
we get that for any $F\in L^p(\R_+; L^q_\tb(\Om))$ problem
\eq{2.7} has, by omitting the exponential weights, a unique solution $(U,\na P)$ such that
$$ (U,\na P)\in L^p(\R_+; \mathcal D(A_{q,\textbf{0}}))
\ti L^p(\R_+; L^q(\Om)),\;\; U_t\in  L^p(\R_+; L^q(\Om)). $$
We shall prove that this solution $(U, \na P)$, furthermore, satisfies
\begin{equation}
\label{E4.33}
(U,\na P)\in
L^p(\R_+; W^{2,q}_\tb(\Om)) \ti L^p(\R_+; L^q_\tb(\Om)), \;
U_t\in L^p(\R_+; L^q_\tb(\Om)).
\end{equation}
Once \eq{4.33} is proved, the (linear) solution operator
$$L^p(\R_+; L^q_\tb(\Om))\ni F\mapsto
(U,\na P)\in L^p(\R_+; \mathcal D(A_{q,\tb}) ) \ti L^p(\R_+; L^q_\tb(\Om))$$
is obviously closed and hence bounded by the closed graph theorem,
thus implying \eq{2.6}.

The proof of \eq{4.33} is based on a cut-off technique using Theorem \ref{T2.3}.
Let $\{\varphi_i\}_{i=0}^{m}$ be the cut-off functions given by \eq{4.15n} and
let $(u^{0},p^{0})$, $(u^i, p^i)$ be defined by \eq{4.7n}.
Then
$(u^i, p^i)$, $i=0,\ldots, m$, satisfies
\begin{equation}\label{E4.17c}
\begin{array}{rcll}
 u^i_t - \Da u^i + \na p^i & =& \tilde{f}^i & \mbox{ in }
                       \R_+\ti\widetilde{\Om}_i,\ek
 \div u^i & = & \tilde{g}^i & \mbox{ in }\R_+\ti\widetilde{\Om}_i,\ek
  u^i(0) & = & 0 & \mbox{ in } \widetilde{\Om}_i,\ek
  u^i & = & 0 & \mbox{ on } \pa \widetilde{\Om}_i,
\end{array}
\end{equation}
where
$$ f^i:=\varphi_iF+(\na\varphi_i)P-(\Da\varphi_i)U-2\na\varphi_i\cdot\na U,
   \quad g^i:=\na\varphi_i\cdot U,\quad i=0,\ldots, m.$$
In view of $g^i\in L^p(\R_+; W^{1,q}_0(\Om_0))$ and
$\int_{\Om_0}g^i\,dx=0$ for $i=0,\ldots,m$, we find
as in the proof of Theorem \ref{T2.4} $w_i\in L^p(\R_+;W^{2,q}_0(\Om_0))$
such that $\div w_i(t)=g^i(t)$ in $\Om_0$ for
almost all $t\in\R_+$, $w_{i,t}\in L^p(\R_+; L^{q}(\Om_0))$  and
\begin{equation}
\label{E4.18}
\begin{array}{l}
\|\na^2 w_i\|_{L^p(\R_+; L^q(\Om_0))}\leq c\|\na g^i\|_{L^p(\R_+; L^q(\Om_0))}
\leq c\|\na U\|_{L^p(\R_+; L^{q}_0(\Om_0))},\ek
\|w_{i,t}\|_{L^p(\R_+; L^q(\Om_0))}\leq c\|g^i_t\|_{L^p(\R_+; (W^{1,q'}(\Om_0))^*)}
\leq c\|U_t\|_{L^p(\R_+; (W^{1,q'}(\Om_0))^*)},
\end{array}
\end{equation}
where $c=c(\Om_0,q)$; here the linearity of the solution operator to the divergence problem
is crucial.
For $i=1\ldots,m$ the extension by $0$ of $w_i$ to $\widetilde{\Om}_i$, say $\tilde{w}_i$,
satisfies
$e^{\be_i x^i_n}\tilde{w}_{i,t}$, $e^{\be_i x^i_n}\na^2 \tilde{w}_i \in L^p(\R_+; L^q(\tilde\Om_i))$
and
\begin{equation}
\label{E4.19}
\begin{array}{l}
\|e^{\be_i x^i_n}\tilde{w}_{i,t},\,
 e^{\be_i x^i_n}\na^2 \tilde{w}_i\|_{L^p(\R_+; L^q(\tilde\Om_i))} \ek
\qquad \leq c\big(\|\na U\|_{L^p(\R_+; L^{q}(\Om_0))}
       + \|U_t\|_{L^p(\R_+; (W^{1,q'}(\Om_0))^*)}\big).
\end{array}
\end{equation}
Moreover, note that $w_i(0,x)=0$ due to $U(0,x)=0$, $g^i(0,x)=0$ for $x\in \Om$.

Now, $v^0:=u^0-w_0$ solves \eq{4.17c} with $f^0$ replaced by $f^0-w_{0,t} +\Da w_0$, and $v^i:=u^i-\tilde{w}_i$, $i=1,\ldots,m$, solves \eq{4.17c} with $\tilde{f}^i$ replaced by $\tilde{f}^i-\tilde{w}_{i,t}$.
Then, by maximal regularity of the Stokes operator in bounded
domains  in view of \eq{4.18} we obtain that
\begin{equation}
\label{E4.20}
\begin{array}{l}
\|v^{0},v^{0}_{t}, \na^2 v^0, \na p^0\|_{L^p(\R_+; L^q(\Om_0))} \ek
\qquad \leq  c\big(\|F,\na U,P\|_{L^p(\R_+; L^{q}(\Om_0))}
 +\|U_t\|_{L^p(\R_+; (W^{1,q'}(\Om_0))^*)} \big),
\end{array}
\end{equation}
and, by Theorem \ref{T2.3} in view of \eq{4.19}, that
\begin{equation}
\label{E4.21}
\begin{array}{l}
\|v^{i},v^{i}_t, \na^2 v^i, \na p^i\|_{L^p(\R_+; L^q_{\be_i}(\R; L^q(\Si^i)))}
\leq  c\big(\|F\|_{L^p(\R_+; L^{q}_{\be_i}(\tilde\Om_i))}\ek
 \qquad + \|\na U, P\|_{L^p(\R_+; L^{q}_0(\Om_0))}
 +\|U_t\|_{L^p(\R_+; (W^{1,q'}(\Om_0))^*)} \big),\; i=1,\ldots,m.
\end{array}
\end{equation}
 Thus, from \eq{4.18}-\eq{4.21} we get that
\begin{equation}
\label{E4.25}
\begin{array}{l}
\|u_0, u^{0}_t, \na^2 u^0, \na p^0\|_{L^p(\R_+; L^q(\Om_0))}\ek
\hspace{2cm}\leq
 c \big(\|F,\na U, P\|_{L^p(\R_+; L^{q}(\Om_0))}
  +\|U_t\|_{L^p(\R_+; (W^{1,q'}(\Om_0))^*)} \big),\ek
 \|u^{i}_{t}, \na^2 u^i, \na p^i\|_{L^p(\R_+;  L^q_{\be_i}(\R; L^q(\Si^i)))}
 \leq
c \big(\|F\|_{L^p(\R_+; L^{q}_{\be_i}(\tilde\Om_i))}\ek
 \hspace{2cm}+\|\na U, P\|_{L^p(\R_+; L^{q}(\Om_0))}
 +\|U_t\|_{L^p(\R_+; (W^{1,q'}(\Om_0))^*)} \big),\; i=1,\ldots,m.
\end{array}
\end{equation}
Since $U=\sum_{i=0}^m u^i$, $P=\sum_{i=0}^m p^i$ in $\Om$,
\eq{4.25} yields \eq{4.33} and
\begin{equation}
\label{E4.32}
\begin{array}{l}
 \|U, U_{t}, \na^2 U, \na P\|_{L^p(\R_+;  L^q_{\tb}(\Om))}\leq
  c \big(\|F\|_{L^p(\R_+; L^{q}_{\tb}(\Om))}\ek
 \hspace{2cm} + \|\na U, P\|_{L^p(\R_+; L^{q}(\Om_0))}+\|U_t\|_{L^p(\R_+; (W^{1,q'}(\Om_0))^*)} \big).
\end{array}
\end{equation}
Note that one may assume without loss of generality that
$\int_{\Om_0}P\,dx=0$. Hence, by Poincar\'e's inequality and the
result of maximal $L^p$-regularity for $1<p<\infty$ of the Stokes
operator in $L^q_\si(\Om)$ without exponential weights (see
\cite[Theorem 1.2]{FR05-4}),
 $\|\na U, P\|_{L^p(\R_+;L^{q}(\Om_0))}+
 \|U_t\|_{L^p(\R_+;(W^{1,q'}(\Om_0))')}$ can be estimated by
$c\|F\|_{L^p(\R_+; L^{q}(\Om))}$ and hence by $c\|F\|_{L^p(\R_+;
L^{q}_{\tb}(\Om))}$ with some constant $c>0$. Consequently, \eq{2.9}
holds true and the Stokes operator $A_{q,\tb}$ in
$L^q_{\tb,\si}(\Om)$
 has maximal $L^p$-regularity for $1<p<\infty$.

The proof of Theorem \ref{T2.5} is complete. \qed
\par\bigskip\noindent
{\bf Acknowledgement:} Part of the work was done during the stay of
the first author in the Institute of Mathematics, AMSS, CAS, China
under the support of 2012 CAS-TWAS Postdoctoral Fellowship, grant
No. 3240267229. He is grateful to Prof. Ping Zhang for inviting him
and CAS (Chinese Academy of Sciences) and TWAS (The World Academy of
Sciences) for financial support.

%

\end{document}